\documentclass[zpreprint,zbstpl]{zeus_paper}

\usepackage[english]{babel}

\chardef\usc=95
\chardef\til=126
\catcode`\@=11 
\DeclareRobustCommand\xdotspace{\futurelet\@let@token\@xdotspace}
\def\@xdotspace{%
  \ifx\@let@token.\else
  \ifx\@let@token\bgroup.\else
  \ifx\@let@token\egroup.\else
  \ifx\@let@token\/.\else
  \ifx\@let@token\ .\else
  \ifx\@let@token~.\else
  \ifx\@let@token!.\else
  \ifx\@let@token,.\else
  \ifx\@let@token:.\else
  \ifx\@let@token;.\else
  \ifx\@let@token?.\else
  \ifx\@let@token/.\else
  \ifx\@let@token'.\else
  \ifx\@let@token).\else
  \ifx\@let@token-.\else
  \ifx\@let@token\@xobeysp.\else
  \ifx\@let@token\space.\else
  \ifx\@let@token\@sptoken.\else
   .\space
   \fi\fi\fi\fi\fi\fi\fi\fi\fi\fi\fi\fi\fi\fi\fi\fi\fi\fi}
\catcode`\@=12 

\newcommand{\stru}[2]{%
   \relax\ifmmode\hbox{\vrule height#1 depth#2 width0pt}%
   \else\vrule height#1 depth#2 width0pt\fi}

\newcommand{\Ronum}[1]{\uppercase\expandafter{\romannumeral#1}}
\newcommand{\ronum}[1]{\expandafter{\romannumeral#1}}
\DeclareRobustCommand{\LaTeXZ}{%
  \LaTeX\kern-.05em4\kern-.1em
  {\raisebox{-0.2ex}{$\scriptstyle\text{ZEUS}$}}\xspace}

\DeclareMathAlphabet{\mathbf}{OT1}{cmr}{bx}{sl}
\newcommand{\eVdist}{\kern-0.06667em}

\newcommand{\Gev}{{\text{Ge}\eVdist\text{V\/}}}

\newcommand{\pb}{\,\text{pb}}
\newcommand{\fb}{\,\text{fb}}

\newcommand{\Tesla}{\,\text{T}}

\newcommand{\slashfrac}[2]{%
  \raisebox{0.5ex}{\ensuremath #1}\kern-0.12em/\kern-0.08em
  \raisebox{-.8ex}{\ensuremath #2}}

\newcommand{\sqr}[3]{%
    {\vcenter{\hrule height.#3ex\hbox{\vrule width.#2ex height#1ex
     \kern#1ex\vrule width.#3ex}\hrule height.#2ex}}}

\catcode`\@=11 
\newcommand{\parenbar}{\mathpalette\p@renb@r}
\def\p@renb@r#1#2{\vbox{%
  \ifx#1\scriptscriptstyle \dimen@.7em\dimen@ii.2em\else
  \ifx#1\scriptstyle \dimen@.8em\dimen@ii.25em\else
  \dimen@1em\dimen@ii.4em\fi\fi \offinterlineskip
  \ialign{\hfill##\hfill\cr
    \vbox{\hrule width\dimen@ii}\cr
    \noalign{\vskip-.3ex}%
    \hbox to\dimen@{$\mathchar300\hfil\mathchar301$}\cr
    \noalign{\vskip-.3ex}%
    $#1#2$\cr}}}
\catcode`\@=12 

\newcommand{\IP}{{\rm I$\kern-0.01667em$P}\xspace}

\newcommand{\F}{{\cal F}}

\mathchardef\qsm=63
\mathchardef\pls=43
\mathchardef\mns=512
\mathchardef\plm=518
\mathchardef\eql=61
\mathchardef\smallleft=300
\mathchardef\smallright=301
\mathchardef\les=316
\mathchardef\gre=318
\mathchardef\leq=532
\mathchardef\grq=533

\catcode`\@=11 
\newcounter{pict@width}
\newcounter{pict@height}
\newlength{\pict@scale}
\newcommand{\psfigadd}[4]{%
\setcounter{pict@width}{1*\ratio{#2+\pict@scale/2}{\pict@scale}}
\setcounter{pict@height}{1*\ratio{#3+\pict@scale/2}{\pict@scale}}
\setlength{\unitlength}{\pict@scale}
\hbox to #2{\hspace{-\fill}\begin{picture}(\thepict@width,\thepict@height)
\put(0,0){\psfig{figure=#1,width=#2,height=#3,clip=}}
\SetScale{0.283466457}
\SetWidth{1.763889}
{#4}
\end{picture}}
}
\newcounter{pict@widthfst}
\newcounter{pict@widthscd}
\newcounter{pict@widthtot}
\newcommand{\psfigaddtwo}[7]{%
\setcounter{pict@widthfst}{1*\ratio{#2+\pict@scale/2}{\pict@scale}}
\setcounter{pict@widthscd}{1*\ratio{#2+#4+\pict@scale/2}{\pict@scale}}
\setcounter{pict@widthtot}{1*\ratio{#2+#4+#6+\pict@scale/2}{\pict@scale}}
\setcounter{pict@height}{1*\ratio{#3+\pict@scale/2}{\pict@scale}}
\setlength{\unitlength}{\pict@scale}
\hbox{\hspace{-\fill}\begin{picture}(\thepict@widthtot,\thepict@height)
\put(0,0){\psfig{figure=#1,width=#2,height=#3,clip=}}
\put(\thepict@widthscd,0){\psfig{figure=#5,width=#6,height=#3,clip=}}
\SetScale{0.283466457}
\SetWidth{1.763889}
{#7}
\end{picture}}
}
\newcommand{\psfigror}[4]{%
\setcounter{pict@width}{1*\ratio{#2+\pict@scale/2}{\pict@scale}}
\setcounter{pict@height}{1*\ratio{#3+\pict@scale/2}{\pict@scale}}
\setlength{\unitlength}{\pict@scale}
\hbox{\begin{picture}(\thepict@width,\thepict@height)
\put(0,\thepict@height){\psfig{figure=#1,width=#3,height=#2,clip=,angle=270}}
\SetScale{0.283466457}
\SetWidth{1.763889}
{#4}
\end{picture}}
}
\newcommand{\psfigrol}[4]{%
\setcounter{pict@width}{1*\ratio{#2+\pict@scale/2}{\pict@scale}}
\setcounter{pict@height}{1*\ratio{#3+\pict@scale/2}{\pict@scale}}
\setlength{\unitlength}{\pict@scale}
\hbox{\begin{picture}(\thepict@width,\thepict@height)
\put(0,0){\psfig{figure=#1,width=#3,height=#2,clip=,angle=90}}
\SetScale{0.283466457}
\SetWidth{1.763889}
{#4}
\end{picture}}
}
\catcode`\@=12 
\newlength\listtextwidth

\catcode`\@=11 
\newlength{\@tabfninsert}
\newlength{\@tabfnwidth}
\newcommand{\tabfootnote}[2]{%
  \setlength{\@tabfninsert}{0.8em}
  \setlength{\@tabfnwidth}{\textwidth}
  \addtolength{\@tabfnwidth}{-\@tabfninsert}
  \addtolength{\@tabfnwidth}{-0.4em}
  \noindent\makebox[\@tabfninsert][r]{\footnotesize$^{#1}$\hfil}\hfill%
  \parbox[t]{\@tabfnwidth}{\footnotesize #2\hfill}}
\catcode`\@=12 

\def\JHEP{JHEP}

\def\phij{\varphi^{\rm jet1}}

\def\qg2{$\q2>125$~\g2}

\def\setjb{d\sigma/d\etjb}
\def\sq2{d\sigma/d\q2}
 
\def\q2{Q^2}

\def\cgh{\cos\gamma_h}
\def\pb1{pb$^{-1}$}
\def\fb1{fb$^{-1}$}

\def\g2{GeV$^2$}

\def\F2g{F_2^{\gamma}}
\def\f2gv{F_2^{\gamma^*}}

\def\rr1{R=1}
\def\r7{R=0.7}
\def\R71{R=0.7\ {\rm and}\ 1}

\def\m3j{M^{\rm 3j}}

\def\kt{k_T}

\def\lq2{\log_{10}(\q2)}

\def\etjb{E^{\rm jet}_{T,{\rm B}}}

\def\etajb{\eta^{\rm jet}_{\rm B}}

\def\etalab{\eta^{\rm jet}_{\rm LAB}}
\def\etlab{E^{\rm jet}_{T,{\rm LAB}}}

\def\ele{e^+e^-}
\def\pp{p\bar p}
\def\qq{q\bar q}

\def\colab#1{#1 Coll.}

\def\z0{Z^0}
\def\mz{M_Z}

\def\as{\alpha_s}
\def\oalphas2{{\cal O}(\alpha\as^2)}

\def\oass{{\cal O}(\as^2)}
\def\oasss{{\cal O}(\as^3)}
\def\asz{\as(\mz)}

\def\asmzn#1#2#3#4#5#6#7{\asz|_{\rm #7} = #1\pm #2\ {\rm (stat.)}\ ^{+#4}_{-#3}\ {\rm (exp.)}\ ^{+#6}_{-#5}\ {\rm (th.)}}

\def\p2{P^2}

\def\mr2{\mu_R^2}
\def\mf2{\mu_F^2}

\def\etal{et al.}

\def\bet0#1#2#3#4#5#6{\beta_0 = #1\pm #2\ {\rm (stat.)}\ ^{+#4}_{-#3}\ {\rm (exp.)}\ ^{+#6}_{-#5}\ {\rm (th.)}}

\def\a34{\alpha_{23}}

\def\etai{\eta^i_{\rm B}}
\def\etaj{\eta^j_{\rm B}}
\def\eti{E^i_{T,{\rm B}}}
\def\etj{E^j_{T,{\rm B}}}
\def\phii{\phi^i_{\rm B}}
\def\phij{\phi^j_{\rm B}}

\def\figdir{./}

\begin{document}
\prepnum{{DESY--10--034}}

\title{Inclusive-jet cross sections in NC DIS at HERA and a comparison
        of the {\boldmath $\kt$}, anti-{\boldmath $\kt$} and SIScone jet
        algorithms}
                    
\author{ZEUS Collaboration}
\date{March 2010}

\abstract{
For the first time, differential inclusive-jet cross sections have
been measured in neutral current deep inelastic $ep$ scattering using
the anti-$\kt$ and SIScone algorithms. The measurements were made for 
boson virtualities $\q2>125$~\g2\ with the ZEUS detector at HERA using
an integrated luminosity of $82$~\pb1\ and the jets were identified in
the Breit frame. The performance and suitability of the jet algorithms
for their use in hadron-like reactions were investigated by comparing
the measurements to those performed with the $\kt$
algorithm. Next-to-leading-order QCD calculations give a good
description of the measurements. Measurements of the ratios of cross
sections using different jet algorithms are also presented; 
the measured ratios are well described by calculations including up to
$\oasss$ terms. Values of $\asz$ were extracted from the data; the results
are compatible with and have similar precision to the value extracted
from the $\kt$ analysis.
}

\makezeustitle

\pagenumbering{Roman}

\begin{center}
{                      \Large  The ZEUS Collaboration              }
\end{center}

{\mbox H.~Abramowicz$^{44, ad}$, }
{\mbox I.~Abt$^{34}$, }
{\mbox L.~Adamczyk$^{13}$, }
{\mbox M.~Adamus$^{53}$, }
{\mbox R.~Aggarwal$^{7}$, }
{\mbox S.~Antonelli$^{4}$, }
{\mbox P.~Antonioli$^{3}$, }
{\mbox A.~Antonov$^{32}$, }
{\mbox M.~Arneodo$^{49}$, }
{\mbox V.~Aushev$^{26, y}$
% ...
}
\newpage

\makebox[3em]{$^{1}$}
\begin{minipage}[t]{14cm}
{\it Argonne National Laboratory, Argonne, Illinois 60439-4815, USA}~$^{A}$

\end{minipage}\\
\makebox[3em]{$^{2}$}
\begin{minipage}[t]{14cm}
{\it Andrews University, Berrien Springs, Michigan 49104-0380, USA}

\end{minipage}\\
\makebox[3em]{$^{3}$}
\begin{minipage}[t]{14cm}
{\it INFN Bologna, Bologna, Italy}~$^{B}$

\end{minipage}\\
\makebox[3em]{$^{4}$}
\begin{minipage}[t]{14cm}
{\it University and INFN Bologna, Bologna, Italy}~$^{B}$

\end{minipage}\\
\makebox[3em]{$^{5}$}
\begin{minipage}[t]{14cm}
{\it Physikalisches Institut der Universit\"at Bonn,
Bonn, Germany}~$^{C}$

\end{minipage}\\
\makebox[3em]{$^{6}$}
\begin{minipage}[t]{14cm}
{\it H.H.~Wills Physics Laboratory, University of Bristol,
Bristol, United Kingdom}~$^{D}$

\end{minipage}\\
\makebox[3em]{$^{7}$}
\begin{minipage}[t]{14cm}
{\it Panjab University, Department of Physics, Chandigarh, India}

\end{minipage}\\
\makebox[3em]{$^{8}$}
\begin{minipage}[t]{14cm}
{\it Calabria University,
Physics Department and INFN, Cosenza, Italy}~$^{B}$

\end{minipage}\\
\makebox[3em]{$^{9}$}
\begin{minipage}[t]{14cm}
{\it Institute for Universe and Elementary Particles, Chonnam National University,\\
Kwangju, South Korea}

\end{minipage}\\
\makebox[3em]{$^{10}$}
\begin{minipage}[t]{14cm}
{\it Jabatan Fizik, Universiti Malaya, 50603 Kuala Lumpur, Malaysia}~$^{E}$

\end{minipage}\\
\makebox[3em]{$^{11}$}
\begin{minipage}[t]{14cm}
{\it Nevis Laboratories, Columbia University, Irvington on Hudson,
New York 10027, USA}~$^{F}$

\end{minipage}\\
\makebox[3em]{$^{12}$}
\begin{minipage}[t]{14cm}
{\it The Henryk Niewodniczanski Institute of Nuclear Physics, Polish Academy of Sciences,\\
Cracow, Poland}~$^{G}$

\end{minipage}\\
\makebox[3em]{$^{13}$}
\begin{minipage}[t]{14cm}
{\it Faculty of Physics and Applied Computer Science, AGH-University of Science and \\
Technology, Cracow, Poland}~$^{H}$

\end{minipage}\\
\makebox[3em]{$^{14}$}
\begin{minipage}[t]{14cm}
{\it Department of Physics, Jagellonian University, Cracow, Poland}

\end{minipage}\\
\makebox[3em]{$^{15}$}
\begin{minipage}[t]{14cm}
{\it Deutsches Elektronen-Synchrotron DESY, Hamburg, Germany}

\end{minipage}\\
\makebox[3em]{$^{16}$}
\begin{minipage}[t]{14cm}
{\it Deutsches Elektronen-Synchrotron DESY, Zeuthen, Germany}

\end{minipage}\\
\makebox[3em]{$^{17}$}
\begin{minipage}[t]{14cm}
{\it INFN Florence, Florence, Italy}~$^{B}$

\end{minipage}\\
\makebox[3em]{$^{18}$}
\begin{minipage}[t]{14cm}
{\it University and INFN Florence, Florence, Italy}~$^{B}$

\end{minipage}\\
\makebox[3em]{$^{19}$}
\begin{minipage}[t]{14cm}
{\it Fakult\"at f\"ur Physik der Universit\"at Freiburg i.Br.,
Freiburg i.Br., Germany}~$^{C}$

\end{minipage}\\
\makebox[3em]{$^{20}$}
\begin{minipage}[t]{14cm}
{\it Department of Physics and Astronomy, University of Glasgow,
Glasgow, United Kingdom}~$^{D}$

\end{minipage}\\
\makebox[3em]{$^{21}$}
\begin{minipage}[t]{14cm}
{\it Department of Engineering in Management and Finance, Univ. of
the Aegean, Chios, Greece}

\end{minipage}\\
\makebox[3em]{$^{22}$}
\begin{minipage}[t]{14cm}
{\it Hamburg University, Institute of Exp. Physics, Hamburg,
Germany}~$^{C}$

\end{minipage}\\
\makebox[3em]{$^{23}$}
\begin{minipage}[t]{14cm}
{\it Imperial College London, High Energy Nuclear Physics Group,
London, United Kingdom}~$^{D}$

\end{minipage}\\
\makebox[3em]{$^{24}$}
\begin{minipage}[t]{14cm}
{\it Institute of Particle and Nuclear Studies, KEK,
Tsukuba, Japan}~$^{I}$

\end{minipage}\\
\makebox[3em]{$^{25}$}
\begin{minipage}[t]{14cm}
{\it Institute of Physics and Technology of Ministry of Education and
Science of Kazakhstan, Almaty, Kazakhstan}

\end{minipage}\\
\makebox[3em]{$^{26}$}
\begin{minipage}[t]{14cm}
{\it Institute for Nuclear Research, National Academy of Sciences, and
Kiev National University, Kiev, Ukraine}

\end{minipage}\\
\makebox[3em]{$^{27}$}
\begin{minipage}[t]{14cm}
{\it Kyungpook National University, Center for High Energy Physics, Daegu,
South Korea}~$^{J}$

\end{minipage}\\
\makebox[3em]{$^{28}$}
\begin{minipage}[t]{14cm}
{\it Institut de Physique Nucl\'{e}aire, Universit\'{e} Catholique de Louvain, Louvain-la-Neuve,\\
Belgium}~$^{K}$

\end{minipage}\\
\makebox[3em]{$^{29}$}
\begin{minipage}[t]{14cm}
{\it Departamento de F\'{\i}sica Te\'orica, Universidad Aut\'onoma
de Madrid, Madrid, Spain}~$^{L}$

\end{minipage}\\
\makebox[3em]{$^{30}$}
\begin{minipage}[t]{14cm}
{\it Department of Physics, McGill University,
Montr\'eal, Qu\'ebec, Canada H3A 2T8}~$^{M}$

\end{minipage}\\
\makebox[3em]{$^{31}$}
\begin{minipage}[t]{14cm}
{\it Meiji Gakuin University, Faculty of General Education,
Yokohama, Japan}~$^{I}$

\end{minipage}\\
\makebox[3em]{$^{32}$}
\begin{minipage}[t]{14cm}
{\it Moscow Engineering Physics Institute, Moscow, Russia}~$^{N}$

\end{minipage}\\
\makebox[3em]{$^{33}$}
\begin{minipage}[t]{14cm}
{\it Moscow State University, Institute of Nuclear Physics,
Moscow, Russia}~$^{O}$

\end{minipage}\\
\makebox[3em]{$^{34}$}
\begin{minipage}[t]{14cm}
{\it Max-Planck-Institut f\"ur Physik, M\"unchen, Germany}

\end{minipage}\\
\makebox[3em]{$^{35}$}
\begin{minipage}[t]{14cm}
{\it NIKHEF and University of Amsterdam, Amsterdam, Netherlands}~$^{P}$

\end{minipage}\\
\makebox[3em]{$^{36}$}
\begin{minipage}[t]{14cm}
{\it Physics Department, Ohio State University,
Columbus, Ohio 43210, USA}~$^{A}$

\end{minipage}\\
\makebox[3em]{$^{37}$}
\begin{minipage}[t]{14cm}
{\it Department of Physics, University of Oxford,
Oxford, United Kingdom}~$^{D}$

\end{minipage}\\
\makebox[3em]{$^{38}$}
\begin{minipage}[t]{14cm}
{\it INFN Padova, Padova, Italy}~$^{B}$

\end{minipage}\\
\makebox[3em]{$^{39}$}
\begin{minipage}[t]{14cm}
{\it Dipartimento di Fisica dell' Universit\`a and INFN,
Padova, Italy}~$^{B}$

\end{minipage}\\
\makebox[3em]{$^{40}$}
\begin{minipage}[t]{14cm}
{\it Department of Physics, Pennsylvania State University, University Park,\\
Pennsylvania 16802, USA}~$^{F}$

\end{minipage}\\
\makebox[3em]{$^{41}$}
\begin{minipage}[t]{14cm}
{\it Polytechnic University, Sagamihara, Japan}~$^{I}$

\end{minipage}\\
\makebox[3em]{$^{42}$}
\begin{minipage}[t]{14cm}
{\it Dipartimento di Fisica, Universit\`a 'La Sapienza' and INFN,
Rome, Italy}~$^{B}$

\end{minipage}\\
\makebox[3em]{$^{43}$}
\begin{minipage}[t]{14cm}
{\it Rutherford Appleton Laboratory, Chilton, Didcot, Oxon,
United Kingdom}~$^{D}$

\end{minipage}\\
\makebox[3em]{$^{44}$}
\begin{minipage}[t]{14cm}
{\it Raymond and Beverly Sackler Faculty of Exact Sciences, School of Physics, \\
Tel Aviv University, Tel Aviv, Israel}~$^{Q}$

\end{minipage}\\
\makebox[3em]{$^{45}$}
\begin{minipage}[t]{14cm}
{\it Department of Physics, Tokyo Institute of Technology,
Tokyo, Japan}~$^{I}$

\end{minipage}\\
\makebox[3em]{$^{46}$}
\begin{minipage}[t]{14cm}
{\it Department of Physics, University of Tokyo,
Tokyo, Japan}~$^{I}$

\end{minipage}\\
\makebox[3em]{$^{47}$}
\begin{minipage}[t]{14cm}
{\it Tokyo Metropolitan University, Department of Physics,
Tokyo, Japan}~$^{I}$

\end{minipage}\\
\makebox[3em]{$^{48}$}
\begin{minipage}[t]{14cm}
{\it Universit\`a di Torino and INFN, Torino, Italy}~$^{B}$

\end{minipage}\\
\makebox[3em]{$^{49}$}
\begin{minipage}[t]{14cm}
{\it Universit\`a del Piemonte Orientale, Novara, and INFN, Torino,
Italy}~$^{B}$

\end{minipage}\\
\makebox[3em]{$^{50}$}
\begin{minipage}[t]{14cm}
{\it Department of Physics, University of Toronto, Toronto, Ontario,
Canada M5S 1A7}~$^{M}$

\end{minipage}\\
\makebox[3em]{$^{51}$}
\begin{minipage}[t]{14cm}
{\it Physics and Astronomy Department, University College London,
London, United Kingdom}~$^{D}$

\end{minipage}\\
\makebox[3em]{$^{52}$}
\begin{minipage}[t]{14cm}
{\it Warsaw University, Institute of Experimental Physics,
Warsaw, Poland}

\end{minipage}\\
\makebox[3em]{$^{53}$}
\begin{minipage}[t]{14cm}
{\it Institute for Nuclear Studies, Warsaw, Poland}

\end{minipage}\\
\makebox[3em]{$^{54}$}
\begin{minipage}[t]{14cm}
{\it Department of Particle Physics, Weizmann Institute, Rehovot,
Israel}~$^{R}$

\end{minipage}\\
\makebox[3em]{$^{55}$}
\begin{minipage}[t]{14cm}
{\it Department of Physics, University of Wisconsin, Madison,
Wisconsin 53706, USA}~$^{A}$

\end{minipage}\\
\makebox[3em]{$^{56}$}
\begin{minipage}[t]{14cm}
{\it Department of Physics, York University, Ontario, Canada M3J
1P3}~$^{M}$

\end{minipage}\\
\vspace{30em} \pagebreak[2]

\makebox[3ex]{$^{ A}$}
\begin{minipage}[t]{14cm}
 supported by the US Department of Energy\
\end{minipage}\\
\makebox[3ex]{$^{ B}$}
\begin{minipage}[t]{14cm}
 supported by the Italian National Institute for Nuclear Physics (INFN) \
\end{minipage}\\
\makebox[3ex]{$^{ C}$}
\begin{minipage}[t]{14cm}
 supported by the German Federal Ministry for Education and Research (BMBF), under
 contract Nos. 05 HZ6PDA, 05 HZ6GUA, 05 HZ6VFA and 05 HZ4KHA\
\end{minipage}\\
\makebox[3ex]{$^{ D}$}
\begin{minipage}[t]{14cm}
 supported by the Science and Technology Facilities Council, UK\
\end{minipage}\\
\makebox[3ex]{$^{ E}$}
\begin{minipage}[t]{14cm}
 supported by an FRGS grant from the Malaysian government\
\end{minipage}\\
\makebox[3ex]{$^{ F}$}
\begin{minipage}[t]{14cm}
 supported by the US National Science Foundation. Any opinion,
 findings and conclusions or recommendations expressed in this material
 are those of the authors and do not necessarily reflect the views of the
 National Science Foundation.\
\end{minipage}\\
\makebox[3ex]{$^{ G}$}
\begin{minipage}[t]{14cm}
 supported by the Polish Ministry of Science and Higher Education as a scientific project No.
 DPN/N188/DESY/2009\
\end{minipage}\\
\makebox[3ex]{$^{ H}$}
\begin{minipage}[t]{14cm}
 supported by the Polish Ministry of Science and Higher Education
 as a scientific project (2009-2010)\
\end{minipage}\\
\makebox[3ex]{$^{ I}$}
\begin{minipage}[t]{14cm}
 supported by the Japanese Ministry of Education, Culture, Sports, Science and Technology
 (MEXT) and its grants for Scientific Research\
\end{minipage}\\
\makebox[3ex]{$^{ J}$}
\begin{minipage}[t]{14cm}
 supported by the Korean Ministry of Education and Korea Science and Engineering
 Foundation\
\end{minipage}\\
\makebox[3ex]{$^{ K}$}
\begin{minipage}[t]{14cm}
 supported by FNRS and its associated funds (IISN and FRIA) and by an Inter-University
 Attraction Poles Programme subsidised by the Belgian Federal Science Policy Office\
\end{minipage}\\
\makebox[3ex]{$^{ L}$}
\begin{minipage}[t]{14cm}
 supported by the Spanish Ministry of Education and Science through funds provided by
 CICYT\
\end{minipage}\\
\makebox[3ex]{$^{ M}$}
\begin{minipage}[t]{14cm}
 supported by the Natural Sciences and Engineering Research Council of Canada (NSERC) \
\end{minipage}\\
\makebox[3ex]{$^{ N}$}
\begin{minipage}[t]{14cm}
 partially supported by the German Federal Ministry for Education and Research (BMBF)\
\end{minipage}\\
\makebox[3ex]{$^{ O}$}
\begin{minipage}[t]{14cm}
 supported by RF Presidential grant N 1456.2008.2 for the leading
 scientific schools and by the Russian Ministry of Education and Science through its
 grant for Scientific Research on High Energy Physics\
\end{minipage}\\
\makebox[3ex]{$^{ P}$}
\begin{minipage}[t]{14cm}
 supported by the Netherlands Foundation for Research on Matter (FOM)\
\end{minipage}\\
\makebox[3ex]{$^{ Q}$}
\begin{minipage}[t]{14cm}
 supported by the Israel Science Foundation\
\end{minipage}\\
\makebox[3ex]{$^{ R}$}
\begin{minipage}[t]{14cm}
 supported in part by the MINERVA Gesellschaft f\"ur Forschung GmbH, the Israel Science
 Foundation (grant No. 293/02-11.2) and the US-Israel Binational Science Foundation \
\end{minipage}\\
\vspace{30em} \pagebreak[2]

\makebox[3ex]{$^{ a}$}
\begin{minipage}[t]{14cm}
also affiliated with University College London,
 United Kingdom\
\end{minipage}\\
\makebox[3ex]{$^{ b}$}
\begin{minipage}[t]{14cm}
now at University of Salerno, Italy\
\end{minipage}\\
\makebox[3ex]{$^{ c}$}
\begin{minipage}[t]{14cm}
now at Queen Mary University of London, United Kingdom\
\end{minipage}\\
\makebox[3ex]{$^{ d}$}
\begin{minipage}[t]{14cm}
also working at Max Planck Institute, Munich, Germany\
\end{minipage}\\
\makebox[3ex]{$^{ e}$}
\begin{minipage}[t]{14cm}
also Senior Alexander von Humboldt Research Fellow at Hamburg University,
 Institute of Experimental Physics, Hamburg, Germany\
\end{minipage}\\
\makebox[3ex]{$^{ f}$}
\begin{minipage}[t]{14cm}
also at Cracow University of Technology, Faculty of Physics,
 Mathemathics and Applied Computer Science, Poland\
\end{minipage}\\
\makebox[3ex]{$^{ g}$}
\begin{minipage}[t]{14cm}
supported by the research grant No. 1 P03B 04529 (2005-2008)\
\end{minipage}\\
\makebox[3ex]{$^{ h}$}
\begin{minipage}[t]{14cm}
now at Rockefeller University, New York, NY
 10065, USA\
\end{minipage}\\
\makebox[3ex]{$^{ i}$}
\begin{minipage}[t]{14cm}
now at DESY group FS-CFEL-1\
\end{minipage}\\
\makebox[3ex]{$^{ j}$}
\begin{minipage}[t]{14cm}
now at DESY group FEB, Hamburg, Germany\
\end{minipage}\\
\makebox[3ex]{$^{ k}$}
\begin{minipage}[t]{14cm}
also at Moscow State University, Russia\
\end{minipage}\\
\makebox[3ex]{$^{ l}$}
\begin{minipage}[t]{14cm}
now at University of Liverpool, United Kingdom\
\end{minipage}\\
\makebox[3ex]{$^{ m}$}
\begin{minipage}[t]{14cm}
on leave of absence at CERN, Geneva, Switzerland\
\end{minipage}\\
\makebox[3ex]{$^{ n}$}
\begin{minipage}[t]{14cm}
now at CERN, Geneva, Switzerland\
\end{minipage}\\
\makebox[3ex]{$^{ o}$}
\begin{minipage}[t]{14cm}
now at Goldman Sachs, London, UK\
\end{minipage}\\
\makebox[3ex]{$^{ p}$}
\begin{minipage}[t]{14cm}
also at Institute of Theoretical and Experimental Physics, Moscow, Russia\
\end{minipage}\\
\makebox[3ex]{$^{ q}$}
\begin{minipage}[t]{14cm}
also at INP, Cracow, Poland\
\end{minipage}\\
\makebox[3ex]{$^{ r}$}
\begin{minipage}[t]{14cm}
also at FPACS, AGH-UST, Cracow, Poland\
\end{minipage}\\
\makebox[3ex]{$^{ s}$}
\begin{minipage}[t]{14cm}
partially supported by Warsaw University, Poland\
\end{minipage}\\
\makebox[3ex]{$^{ t}$}
\begin{minipage}[t]{14cm}
partially supported by Moscow State University, Russia\
\end{minipage}\\
\makebox[3ex]{$^{ u}$}
\begin{minipage}[t]{14cm}
also affiliated with DESY, Germany\
\end{minipage}\\
\makebox[3ex]{$^{ v}$}
\begin{minipage}[t]{14cm}
now at Japan Synchrotron Radiation Research Institute (JASRI), Hyogo, Japan\
\end{minipage}\\
\makebox[3ex]{$^{ w}$}
\begin{minipage}[t]{14cm}
also at University of Tokyo, Japan\
\end{minipage}\\
\makebox[3ex]{$^{ x}$}
\begin{minipage}[t]{14cm}
now at Kobe University, Japan\
\end{minipage}\\
\makebox[3ex]{$^{ y}$}
\begin{minipage}[t]{14cm}
supported by DESY, Germany\
\end{minipage}\\
\makebox[3ex]{$^{ z}$}
\begin{minipage}[t]{14cm}
partially supported by Russian Foundation for Basic Research grant
 \mbox{No. 05-02-39028-NSFC-a}\
\end{minipage}\\
\makebox[3ex]{$^{\dagger}$}
\begin{minipage}[t]{14cm}
 deceased \
\end{minipage}\\
\makebox[3ex]{$^{aa}$}
\begin{minipage}[t]{14cm}
STFC Advanced Fellow\
\end{minipage}\\
\makebox[3ex]{$^{ab}$}
\begin{minipage}[t]{14cm}
nee Korcsak-Gorzo\
\end{minipage}\\
\makebox[3ex]{$^{ac}$}
\begin{minipage}[t]{14cm}
This material was based on work supported by the
 National Science Foundation, while working at the Foundation.\
\end{minipage}\\
\makebox[3ex]{$^{ad}$}
\begin{minipage}[t]{14cm}
also at Max Planck Institute, Munich, Germany, Alexander von Humboldt
 Research Award\
\end{minipage}\\
\makebox[3ex]{$^{ae}$}
\begin{minipage}[t]{14cm}
now at Nihon Institute of Medical Science, Japan\
\end{minipage}\\
\makebox[3ex]{$^{af}$}
\begin{minipage}[t]{14cm}
now at SunMelx Co. Ltd., Tokyo, Japan\
\end{minipage}\\
\makebox[3ex]{$^{ag}$}
\begin{minipage}[t]{14cm}
now at Osaka University, Osaka, Japan\
\end{minipage}\\
\makebox[3ex]{$^{ah}$}
\begin{minipage}[t]{14cm}
now at University of Bonn, Germany\
\end{minipage}\\
\makebox[3ex]{$^{ai}$}
\begin{minipage}[t]{14cm}
also at \L\'{o}d\'{z} University, Poland\
\end{minipage}\\
\makebox[3ex]{$^{aj}$}
\begin{minipage}[t]{14cm}
member of \L\'{o}d\'{z} University, Poland\
\end{minipage}\\
\makebox[3ex]{$^{ak}$}
\begin{minipage}[t]{14cm}
now at Lund University, Lund, Sweden\
\end{minipage}\\
\makebox[3ex]{$^{al}$}
\begin{minipage}[t]{14cm}
also at University of Podlasie, Siedlce, Poland\
\end{minipage}\\

\newpage

\pagenumbering{arabic} 
\pagestyle{plain}

\section{Introduction}
\label{intro}

Jet production at high transverse energies in neutral-current (NC)
deep inelastic $ep$ scattering (DIS) at HERA provides unique tests of
perturbative QCD (pQCD) in a cleaner hadron-like enviroment than that
encountered in hadron-hadron collisions. In NC DIS, the jet search is
performed in the Breit frame~\cite{bookfeynam:1972,*zfp:c2:237} in
which, at leading order (LO) in the strong coupling constant, $\as$,
the boson-gluon-fusion ($V^{*} g\rightarrow \qq$, with $V=\gamma, \z0$) 
and QCD-Compton ($V^{*} q \rightarrow qg$) processes give rise to two
hard jets with opposite transverse momenta.

In previous publications, the observables used to test pQCD included
inclusive-jet~\cite{pl:b547:164,pl:b551:226,np:b765:1,pl:b649:12,
epj:c19:289,pl:b542:193,pl:b653:134,epj:c65:363},
dijet~\cite{pl:b507:70,epj:c23:13,np:b765:1,epj:c19:289,epj:c65:363}
and multijet~\cite{epj:c44:183,desy-08-100,pl:b515:17,epj:c65:363}
cross sections and the internal structure of
jets~\cite{pl:b558:41,np:b700:3,epj:c63:527,np:b545:3} defined using
the $\kt$ cluster algorithm~\cite{np:b406:187} in the longitudinally
invariant inclusive mode~\cite{pr:d48:3160}. These studies
demonstrated that this jet algorithm results in the smallest
uncertainties in the reconstruction of jets in $ep$ collisions.
The $\kt$ algorithm is well suited for $ep$
collisions and yields infrared- and collinear-safe cross sections at
any order of pQCD. However, it might not be best suited to reconstruct
jets in hadron-hadron collisions, such as those at the LHC. In order
to optimise the reconstruction of jet observables in such
environments, new algorithms were recently developed, like the
anti-$\kt$~\cite{jhep:04:063}, a recombination-type jet algorithm, and
the ``Seedless Infrared-Safe'' cone (SIScone)~\cite{jhep:05:086}
algorithms. The measurements of jet cross sections in NC DIS using
these algorithms provide additional tests of QCD and also a test of
their performance with data in a well understood hadron-induced
reaction.

This letter presents measurements of differential inclusive-jet
cross sections as a function of the jet transverse energy in the Breit
frame, $\etjb$, and the virtuality of the exchanged boson, $\q2$,
based on the anti-$\kt$ and SIScone jet algorithms. Differential
inclusive-jet cross sections as functions of $\etjb$ in different
regions of $\q2$ are also presented. The analysis is
based on the same data sample which was used in recent
publications~\cite{pl:b649:12,np:b765:1} to make cross-section
measurements with the $\kt$ jet algorithm. The results are
compared with next-to-leading-order (NLO) QCD calculations using recent
parameterisations of the parton distribution functions (PDFs) of the
proton~\cite{pr:d67:012007,jhep:0207:012,*jhep:0310:046,epj:c63:189}
and with previous measurements based on the $\kt$
algorithm. Furthermore, measurements of the ratios of inclusive-jet
cross sections using different jet algorithms are presented and
compared to pQCD calculations including up to $\oasss$ terms. In
addition, new determinations of $\asz$ were obtained so as to quantify
the performance of the anti-$\kt$ and SIScone algorithms in comparison
with that of the $\kt$ analysis.

\section{Jet algorithms}

The study of jet production in hadronic reactions has been well
established as a testing ground of pQCD. Such tests rely on a
trustworthy reconstruction of the topology of the final-state
partons. This is best done by using jet algorithms with analogous
implementation in experiment and theory. On the experimental side, the
results of the application of a jet algorithm should not depend
significantly on the presence of soft particles or particles which
undergo strong decays. On the theoretical side, they should be
infrared and collinear safe, so that order-by-order pQCD calculations
can be performed. A close correspondence between the jets and the
final-state partons, so that hadronisation corrections are small, and
suppression of beam-remnant jet contributions, when necessary, are
also required.

There are two different methods to reconstruct jets from
the final-state particles: cluster- and cone-type algorithms. 
Cluster algorithms, such as JADE~\cite{zp:c33:23,*pl:b213:235}
or $\kt$, are based on successive recombinations
of particles. They are usually applied in $\ele$ experiments, where
the initial state is governed by QED and the final-state hadrons arise
uniquely from the short-distance interaction so that all hadrons observed
should be associated with the hard process and thus clustered. Cone
algorithms, such as the iterative cone algorithm~\cite{pl:b123:115},
are based on a maximisation of the energy density within a cone of
fixed size and are usually applied in hadron collisions, where the initial
state consists of coloured partons. The initial parton carries only a
fraction of the momentum of the parent hadron, and the
spectator partons lead to the presence of remnant jets and the
underlying event, which leads to the production of soft hadrons not
correlated with the hard interaction. Thus, not all final-state
particles would be associated to jets.

In NC DIS, jets are usually defined using the transverse-energy
flow in the pseudorapidity ($\eta_{\rm B}$)$-$azimuth ($\phi_{\rm B}$)
plane of the Breit frame. The procedure to reconstruct jets with the
$\kt$ algorithm from an initial list of objects (e.g. final-state partons,
final-state hadrons or energy deposits in the calorimeter)
is described below in some detail. In the following discussion, $\eti$ 
denotes the transverse energy, $\etai$ the pseudorapidity and
$\phii$ the azimuthal angle of object $i$ in the Breit frame. For
each pair of objects, the quantity

\begin{equation}
d_{ij}={\rm min}((\eti)^2,(\etj)^2)\cdot[(\etai-\etaj)^2+(\phii-\phij)^2]/R^2
\end{equation}
is calculated. For each individual object, the distance to the beam,
$d_i=(\eti)^2$, is also calculated. If, of all the values
$\{d_{ij},d_i\}$, $d_{kl}$ is the smallest, then objects $k$ and $l$
are combined into a single new object. If, however, $d_k$ is the
smallest, then object $k$ is considered a jet and removed from the
sample. The procedure is repeated until all objects are assigned to jets.

The $\kt$ algorithm is infrared and collinear safe to all orders in
pQCD. As a result of the distance between objects defined in Eq.~(1),
the jets thus obtained have irregular shapes~\cite{jhep:0804:005}, in
contrast to jets defined using a cone algorithm. From an experimental
point of view, it may be desirable for jets to have a well-defined
shape, e.g. to calibrate the jet-energy scale or to estimate the
underlying-event contribution. On the other hand, cone algorithms
produce jets that have an approximate circular shape in the
pseudorapidity-azimuth plane. However, most implementations of
cone algorithms rely on ``seeds'' to start the search for stable
cones, which results in unsafe infrared and collinear behaviour beyond
a given order in pQCD.

Recently, new jet algorithms have been proposed which produce
jets with an approximate circular shape but maintain the
infrared and collinear safety to all orders in pQCD. Two approaches have
been followed. The SIScone jet algorithm has been developed to
overcome the seed problem. A new recombination-type algorithm, the
anti-$\kt$, has been devised which, after a modification of the
distance as defined in Eq.~(1), yields circular jets.

The SIScone algorithm consists of two steps. First, for a given set of
initial objects, all stable cones are identified; cones are classified
as stable by the coincidence of the cone axis with that defined by 
the total momentum of the objects contained in the given cone of
radius $R$ in the $\eta_{\rm B}-\phi_{\rm B}$ plane of the Breit
frame. In this procedure, no seed is used. Stable cones are then
discarded if their transverse momentum is below a given threshold,
$p_{t,{\rm min}}$. For each selected stable cone, the scalar sum of
the transverse momentum of the objects associated to it,
$\tilde{p}_t$, is defined. Second, overlapping cones are identified
and subsequently split or merged according to the following
procedure. Two cones are merged if the scalar sum of the transverse 
momentum of the objects shared by the two cones exceeds a certain
fraction $f$ of the lowest-$\tilde{p}_t$ cone; otherwise, two
different cones are considered and the common objects are assigned to
the nearest cone. This jet algorithm is infrared and collinear safe to
all orders in pQCD.

The $\kt$ and anti-$\kt$ algorithms are identical except for a
modified distance measure,

\begin{equation}
d_{ij}={\rm min}((\eti)^{-2},(\etj)^{-2})\cdot[(\etai-\etaj)^2+(\phii-\phij)^2]/R^2,
\end{equation}

and the distance to the beam, which is defined as $d_i=(\eti)^{-2}$.
This procedure preserves the infrared and collinear safety of the
$\kt$ algorithm and, in addition, gives rise to circular jets.

For the measurements presented in this letter, the parameter $R$ was
set to unity and the jet variables were defined according to the
Snowmass convention~\cite{proc:snowmass:1990:134}. In the application
of the SIScone algorithm, the fraction $f$ was set to $0.75$ and the
transverse momentum threshold $p_{t,{\rm min}}$ was set to zero.

\section{Experimental set-up}
A detailed description of the ZEUS detector can be found
elsewhere~\cite{pl:b293:465,zeus:1993:bluebook}. A brief outline of
the components that are most relevant for this analysis is given below.

Charged particles were tracked in the central tracking detector
(CTD)~\cite{nim:a279:290,*npps:b32:181,*nim:a338:254}, which operated
in a magnetic field of $1.43\Tesla$ provided by a thin superconducting
solenoid. The CTD consisted of $72$~cylindrical drift-chamber
layers, organised in nine superlayers covering the
polar-angle\footnote{The ZEUS coordinate system is a right-handed
  Cartesian system, with the $Z$ axis pointing in the proton beam
  direction, referred to as the ``forward direction'', and the $X$
  axis pointing towards the centre of HERA. The coordinate origin
  is at the nominal interaction point.}
region \mbox{$15^\circ<\theta<164^\circ$}. The transverse-momentum
resolution for full-length tracks can be parameterised as
$\sigma(p_T)/p_T=0.0058p_T\oplus0.0065\oplus0.0014/p_T$, with $p_T$ in
$\Gev$. The tracking system was used to measure the interaction vertex
with a typical resolution along (transverse to) the beam direction of
$0.4$~($0.1$)~cm and to cross-check the energy scale of the calorimeter.

The high-resolution uranium--scintillator calorimeter
(CAL)~\cite{nim:a309:77,*nim:a309:101,*nim:a321:356,*nim:a336:23} covered
$99.7\%$ of the total solid angle and consisted of three parts: the
forward (FCAL), the barrel (BCAL) and the rear (RCAL) calorimeters. 
Each part was subdivided transversely into towers and longitudinally
into one electromagnetic section and either one (in RCAL) or two (in
BCAL and FCAL) hadronic sections. The smallest subdivision of the
calorimeter was called a cell. Under test-beam conditions, the CAL
single-particle relative energy resolutions were
$\sigma(E)/E=0.18/\sqrt E$ for electrons and 
$\sigma(E)/E=0.35/\sqrt E$ for hadrons, with $E$ in GeV.

The luminosity was measured from the rate of the bremsstrahlung process
$ep\rightarrow e\gamma p$. The resulting small-angle energetic photons
were measured by the luminosity
monitor~\cite{desy-92-066,*zfp:c63:391,*acpp:b32:2025}, a
lead--scintillator calorimeter placed in the HERA tunnel at $Z=-107$ m.

\section{Data selection}
The data were collected during the running period 1998--2000, when HERA
operated with protons of energy $E_p=920$~GeV and
electrons\footnote{Here and in the following, the term ``electron'' 
  denotes generically both the electron ($e^-$) and the positron
  ($e^+$).} of energy $E_e=27.5$~GeV, and
correspond to an integrated luminosity of $81.7\pm 1.8$~\pb1, of which
$16.7$~\pb1\ ($65.0$~\pb1) was for $e^-p$ ($e^+p$) collisions.

Neutral current DIS events were selected using the same
criteria as reported in recent
publications~\cite{pl:b649:12,np:b765:1}. The main steps are briefly listed
below.
 
The scattered-electron candidate was identified from the pattern of
energy deposits in the CAL~\cite{nim:a365:508,*nim:a391:360}. The
energy ($E_e^{\prime}$) and polar angle of the
electron candidate were determined from the CAL measurements. 
The $\q2$ variable was reconstructed using the double-angle 
method~\cite{proc:hera:1991:23,*proc:hera:1991:43}.
The angle $\gamma_h$, which corresponds to the angle of the scattered
quark in the quark-parton model, was reconstructed using the hadronic
final state~\cite{proc:hera:1991:23,*proc:hera:1991:43}.

The main requirements imposed on the data sample were:
an electron candidate with $E_{e}^{\prime}>10$~GeV;
a vertex position along the beam axis in the range $|Z|<34$~cm; 
$38<(E-P_Z)<65$~GeV, where $E$ is the total energy as measured
by the CAL, $E=\sum_iE_i$, and $P_Z$ is the $Z$ component of the
vector ${\bf P}=\sum_i {E_i} {\bf\hat r}_i$; in both cases the sum runs
over all CAL cells, $E_i$ is the energy of the CAL cell $i$
and ${\bf \hat r}_i$ is a unit vector along the line joining the
reconstructed vertex and the geometric centre of the cell $i$;
$\q2>125$~\g2; and $|\cgh|<0.65$. In this selected sample,
contamination from non-$ep$ interactions and other physics processes
was negligible.

The anti-$\kt$ and SIScone algorithms\footnote{The implementations of the
anti-$\kt$ and SIScone algorithms in the {\sc Fastjet}
package~\cite{pl:b641:57}, version 2.4.1, were used.} were both used
to reconstruct jets in the hadronic final state both in
data and in Monte Carlo (MC) simulated events (see
Section~\ref{mc}). In the data, the algorithms were applied to the
energy deposits measured in the CAL cells after excluding those
associated with the scattered-electron candidate.
After reconstructing the jet variables in the Breit frame, the
massless four-momenta were boosted into the laboratory frame, where
the transverse energy~($\etlab$) and the pseudorapidity~($\etalab$) of
each jet were calculated. Energy
corrections~\cite{pl:b547:164,pl:b531:9,pl:b558:41} were then applied
to the jets in the laboratory frame,
separately for the anti-$\kt$ and SIScone jet samples. In addition,
events were removed from the samples if any of the jets was in the
backward region of the detector ($\etalab<-2$) and jets were not
included in the final samples if $\etlab<2.5$~GeV.

Only events with at least one jet in the pseudorapidity range
$-2<\etajb<1.5$ were kept for further analysis. The final data samples
with at least one jet satisfying $\etjb>8$~GeV contained $18847$ events
(12575 one-jet, 6126 two-jet, 145 three-jet and one four-jet) in the
anti-$\kt$ sample and $19486$ events (13081 one-jet, 6252 two-jet, 152
three-jet and one four-jet) in the SIScone sample.

\section{Monte Carlo simulation}
\label{mc}
Samples of events were generated to determine the response of the
detector to jets of hadrons and to compute the correction factors
necessary to obtain the hadron-level jet cross sections. The hadron
level is defined by those hadrons with lifetime $\tau\geq 10$~ps. The
generated events were passed through the {\sc
  Geant}~3.13-based~\cite{tech:cern-dd-ee-84-1} ZEUS detector- and
trigger-simulation programs~\cite{zeus:1993:bluebook}. They were
reconstructed and analysed by the same program chain as the data.
 
Neutral current DIS events including electroweak radiative effects
were simulated using the 
{\sc Heracles}~4.6.1~\cite{cpc:69:155,*spi:www:heracles} program with
the {\sc Djangoh}~1.1~\cite{cpc:81:381,*spi:www:djangoh11} interface
to the QCD programs. The QCD cascade was simulated using the
colour-dipole model
(CDM)~\cite{pl:b165:147,*pl:b175:453,*np:b306:746,*zfp:c43:625} 
including the LO QCD diagrams as implemented in
{\sc Ariadne}~4.08~\cite{cpc:71:15,*zfp:c65:285} and, alternatively, 
with the MEPS model of {\sc Lepto} 6.5~\cite{cpc:101:108}. 
The CTEQ5D~\cite{epj:c12:375} parameterisations of the proton PDFs were
used for these simulations. Fragmentation into hadrons was performed
using the Lund string model~\cite{prep:97:31} as implemented in 
{\sc Jetset}~\cite{cpc:82:74,*cpc:135:238,cpc:39:347,*cpc:43:367}.
 
The jet search was performed on the simulated events using the energy
measured in the CAL cells in the same way as for the data. 
The same jet algorithms were also applied to the final-state particles
(hadron level) and to the partons available after the parton shower
(parton level). Additional MC samples were used to correct the measured
cross sections for QED radiative effects and the running of 
$\alpha_{\rm em}$.

\section{Acceptance corrections and experimental uncertainties}
\label{expunc}

To measure the cross sections, the $\etjb$ and $\q2$ distributions in
the data were corrected for detector effects using bin-by-bin
correction factors determined with the MC samples. These correction
factors took into account the efficiency of the selection criteria and
the purity and efficiency of the jet reconstruction. For this approach
to be valid, the uncorrected distributions of the data must be well
described by the MC simulations at the detector level. This condition
was satisfied by both the {\sc Ariadne} and {\sc Lepto}-MEPS MC
samples. The average between the acceptance-correction values obtained
with {\sc Ariadne} and {\sc Lepto}-MEPS was used to correct the data
to the hadron level. The deviations in the results obtained by using
either {\sc Ariadne} or {\sc Lepto}-MEPS to correct the data from
their average were taken to represent systematic uncertainties of the
effect of the QCD-cascade model in the corrections (see below). The
acceptance-correction factors differed from unity by typically less
than $10\%$. 

The following sources of systematic uncertainty were considered for
the measured cross sections:
\begin{itemize}
  \item the uncertainty in the absolute energy scale of the jets was
    estimated to be $\pm 1\%$ for $\etlab>10$~GeV and $\pm 3\%$ for
    lower $\etlab$
    values~\cite{pl:b531:9,epj:c23:615,proc:calor:2002:767}. The
    resulting uncertainty on the cross sections was about $\pm 5\%$;
  \item the differences in the results obtained by using either
    {\sc Ariadne} or {\sc Lepto}-MEPS to correct the data for detector
    effects were taken to represent systematic uncertainties. The
    resulting uncertainty on the cross sections was typically below
    $\pm 3\%$;
  \item the uncertainty due to the selection cuts was estimated by
    varying the values of the cuts within the resolution of each
    variable; the effect on the cross sections was typically below
    $\pm 3\%$;
  \item the uncertainty on the reconstruction of the boost to the
    Breit frame was estimated by using the direction of the track
    associated with the scattered electron instead of that derived from
    its impact position in the CAL. The effect on the cross sections
    was typically below $\pm 1\%$;
  \item the uncertainty in the absolute energy scale of the electron
    candidate was estimated to be $\pm 1\%$~\cite{epj:c21:443}. The
    resulting uncertainty on the cross sections was below $\pm 1\%$;
  \item the uncertainty in the cross sections due to that in the
    simulation of the trigger was negligible.
\end{itemize}
The systematic uncertainties were similar for both jet algorithms.
Those not associated with the absolute energy scale of the jets were
added in quadrature to the statistical uncertainties and are shown in
the figures as error bars. The uncertainty due to the absolute energy
scale of the jets is shown separately as a shaded band due to the large
bin-to-bin correlation. In addition, there was an overall
normalisation uncertainty of $2.2\%$ from the luminosity
determination, which is not included in the figures.

\section{QCD calculations and theoretical uncertainties}
\label{thun}
Next-to-leading-order ($\oass$) QCD calculations were compared to the
measured inclusive-jet cross sections. The calculations were obtained
using the program {\sc Disent}~\cite{np:b485:291}, in the
$\overline{\rm MS}$ renormalisation and factorisation schemes using a
generalised version~\cite{np:b485:291} of the subtraction
method~\cite{np:b178:421}. The number of flavours was set to five and 
the renormalisation ($\mu_R$) and factorisation ($\mu_F$) scales were
chosen to be $\mu_R=\etjb$ and $\mu_F=Q$, respectively. The strong
coupling constant was calculated at two loops with
$\Lambda^{(5)}_{\overline{\rm MS}}=226$~MeV, corresponding to
$\asz=0.118$. The calculations were performed using the
ZEUS-S~\cite{pr:d67:012007} parameterisations of the proton
PDFs\footnote{The LHAPDF package~\cite{hep-ph:0508110}, version 5.7.1,
  was used.}. In {\sc Disent}, the value of $\alpha_{\rm em}$ was fixed
to $1/137$. The anti-$\kt$ and SIScone algorithms were also applied to
the partons in the events generated by {\sc Disent} in order to
compute the jet cross-section predictions.

\begin{figure}[p]
\vfill
\setlength{\unitlength}{1.0cm}
\begin{picture} (18.0,10.0)
\put (0.0,0.0){\centerline{\epsfig{figure=\figdir 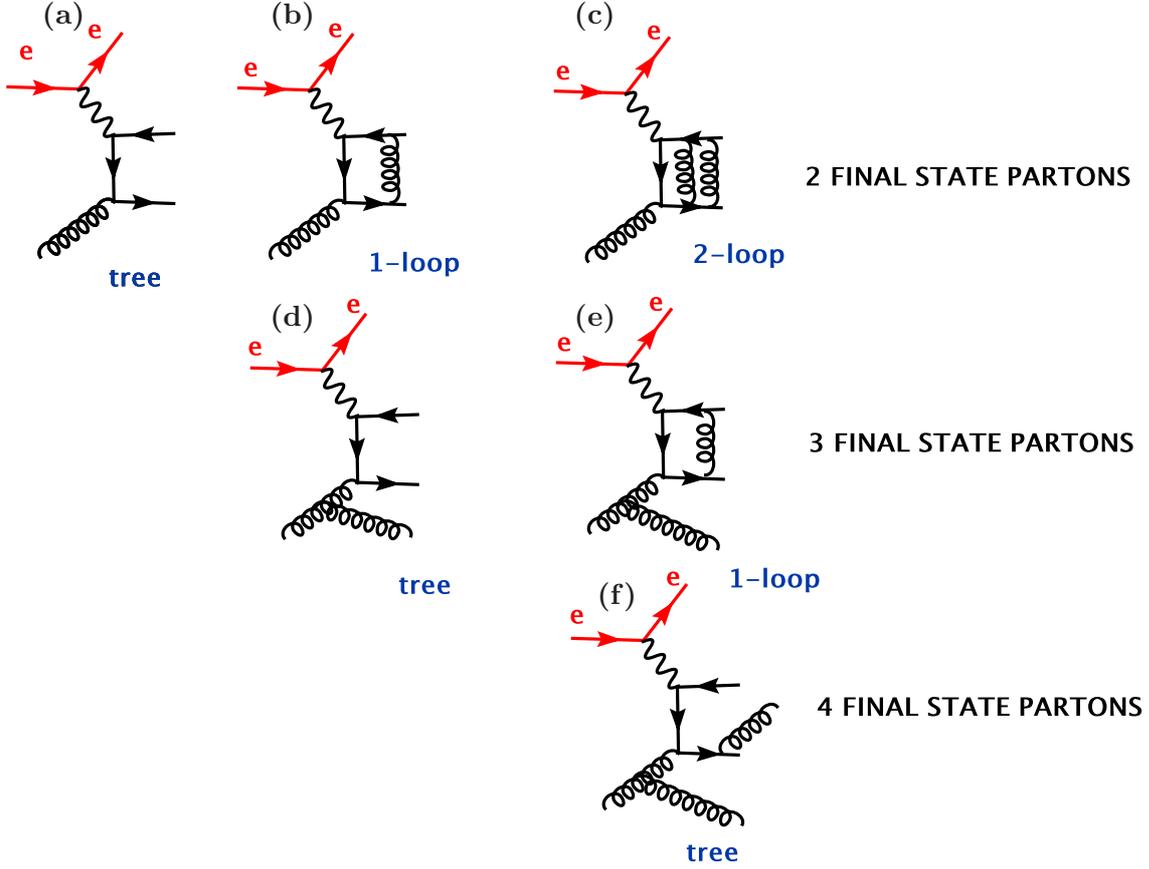,width=15cm}}}
\put (1.0,11.2){\bf\small (a)}
\put (4.0,11.2){\bf\small (b)}
\put (8.0,11.2){\bf\small (c)}
\put (4.0,7.2){\bf\small (d)}
\put (8.0,7.2){\bf\small (e)}
\put (8.3,3.5){\bf\small (f)}
\end{picture}
\caption
{\it 
Examples of Feynman diagrams contributing to inclusive-jet production
in the Breit frame up to $\oasss$. The calculations of inclusive-jet
cross sections up to $\oass$ include the (a) leading-order diagrams and
(b) virtual and (d) real corrections. The lowest-order diagrams that
contribute to the cross-section difference between the anti-$\kt$ and
$\kt$ algorithms are of type (f). The calculations of the
cross-section difference between the SIScone and $\kt$ algorithms
include diagrams of type (d), (e) and (f).
}
\label{fig0}
\vfill
\end{figure}

Measurements are also presented of the ratio of inclusive-jet cross
sections based on different jet algorithms. Although the calculations
for inclusive-jet cross sections can be made at present only up to
$\oass$ (see Fig.~\ref{fig0}), the differences\footnote{The
  differences of cross sections were evaluated on an event-by-event
  basis.} of cross sections using different jet algorithms can be
predicted up to $\oasss$ using the program {\sc
  Nlojet++}~\cite{prl:87:082001}. The same parameter settings as
defined above were used for the {\sc Nlojet++} predictions. The
predicted ratio $(d\sigma_{{\rm anti-}\kt}/dX)/(d\sigma_{\kt}/dX)$,
where $X=\q2$ or $\etjb$, was calculated as

\begin{equation}
\frac{d\sigma_{{\rm anti-}\kt}/dX}{d\sigma_{\kt}/dX}=
1+\frac{d\sigma_{{\rm anti-}\kt}/dX - d\sigma_{\kt}/dX}{d\sigma_{\kt}/dX}\simeq
1+\frac{C\cdot \as^3}{A\cdot \as + B\cdot \as^2},
\end{equation}
since differences between the anti-$\kt$ and $\kt$ algorithms appear
first for final states with four partons and were evaluated using the
tree-level $eq\rightarrow eqggg$, $eq\rightarrow eqg\qq$,
$eg\rightarrow e\qq\qq$ and $eg\rightarrow egg\qq$ subprocesses (see
Fig.~\ref{fig0}). The predicted ratio $(d\sigma_{\rm
  SIScone}/dX)/(d\sigma_{\kt}/dX)$ was calculated as 

\begin{equation}
\frac{d\sigma_{\rm SIScone}/dX}{d\sigma_{\kt}/dX}=
1+\frac{d\sigma_{\rm SIScone}/dX - d\sigma_{\kt}/dX}{d\sigma_{\kt}/dX}\simeq
1+\frac{D\cdot \as^2 + E\cdot \as^3}{A\cdot \as + B\cdot \as^2},
\end{equation}
since differences between the SIScone and $\kt$ algorithms appear
first for final states with three partons and were evaluated using the
tree-level three-parton and four-parton subprocesses and the one-loop
three-parton configurations (see Fig.~\ref{fig0}). Equation~(4) also
applies to the predicted ratio 
$(d\sigma_{{\rm anti-}\kt}/dX)/(d\sigma_{\rm SIScone}/dX)$.

Since the measurements refer to jets of hadrons, whereas the NLO QCD
calculations refer to jets of partons, the predictions were corrected
to the hadron level using the MC models. The multiplicative correction
factor ($C_{\rm had}$) was defined as the ratio of the cross section
for jets of hadrons over that for jets of partons, estimated by using
the MC programs described in Section~\ref{mc}. The mean of the ratios 
obtained with {\sc Ariadne} and {\sc Lepto-MEPS} was taken as the value
of $C_{\rm had}$. The value of $C_{\rm had}$ for the inclusive-jet
cross sections differs from unity by less than
$5\%$~\cite{pl:b649:12}, $6\%$ and $11\%$ for the $\kt$, anti-$\kt$
and SIScone jet algorithms, respectively, in the region $\q2\geq 500$~\g2.

Neither {\sc Disent} nor {\sc Nlojet++} include the contribution from
$\z0$ exchange; MC simulated events with and without $\z0$ exchange
were used to include this effect in the pQCD predictions. In the
following, pQCD calculations refer to the fully corrected predictions.

Several sources of uncertainty in the theoretical predictions for the
inclusive-jet cross sections were considered:
\begin{itemize}
  \item the uncertainty on the NLO QCD calculations due to terms
    beyond NLO, estimated by varying $\mu_R$ between $\etjb/2$ and
    $2\etjb$, was below $\pm 7\ (\pm 10)\%$ at low $\q2$ and low
    $\etjb$ and decreased to less than $\pm 5\ (\pm 7)\%$ for $\q2 >
    250$~\g2\ for the $\kt$~\cite{pl:b649:12} and anti-$\kt$ (SIScone)
    algorithms;
  \item the uncertainty on the NLO QCD calculations due to those on the 
    proton PDFs was estimated by repeating the calculations using 22
    additional sets from the ZEUS-S analysis, which takes into
    account the statistical and correlated systematic experimental
    uncertainties of each data set used in the determination of the
    proton PDFs. The resulting uncertainty in the cross sections for
    all three jet algorithms was below $\pm 3\%$, except in the
    high-$\etjb$ region where it reached $\pm 4\%$;
  \item the uncertainty on the NLO QCD calculations due to that on
    $\asz$ was estimated by repeating the calculations using two
    additional sets of proton PDFs, for which different values of
    $\asz$ were assumed in the fits. The difference between
    the calculations using these various sets was scaled by a factor
    to reflect the uncertainty on the current world average of
    $\as$~\cite{jp:g26:r27}. The resulting uncertainty in the
    cross sections was below $\pm 2\%$ for all three jet algorithms;
  \item the uncertainty from the modelling of the QCD cascade was
    assumed to be half the difference between the hadronisation
    corrections obtained using the {\sc Ariadne} and {\sc Lepto-MEPS}
    models. The resulting uncertainty on the cross sections was less
    than $\pm 1.4\%$~\cite{pl:b649:12}, $\pm 1.7\%$ and $\pm 2.3\%$
    for the $\kt$, anti-$\kt$ and SIScone algorithms, respectively;
  \item the uncertainty of the calculations due to the choice of
    $\mu_F$ was estimated by repeating the calculations with
    $\mu_F=Q/2$ and $2Q$. The effect was negligible.
\end{itemize}
 
The total theoretical uncertainty was obtained by adding in quadrature
the individual uncertainties listed above. As a function of $\q2$, the
total theoretical uncertainty varies in the range $3-7\%$ ($3-10\%$)
for the anti-$\kt$ (SIScone) algorithm; as a function of $\etjb$, the
range of variation is $5-6\%$ ($6-8\%$) for the anti-$\kt$ (SIScone)
algorithm.

It is concluded that the NLO QCD predictions for the inclusive-jet
cross sections are of similar precision for the $\kt$ and anti-$\kt$
algorithms and somewhat less precise for the SIScone. A more detailed
comparison of the theoretical uncertainties is presented in
Section~\ref{results}.

\section{Results}
\label{results}

\subsection{Inclusive-jet differential cross sections with different
  jet algorithms}
The inclusive-jet differential cross sections were measured in the
kinematic region $\q2>125$~\g2\ and $|\cgh|<0.65$. The jets were
reconstructed using either the anti-$\kt$ or the SIScone jet
algorithms and the cross sections refer to jets with $\etjb>8$~GeV and
$-2<\etajb<1.5$. These cross sections were corrected for detector and
QED radiative effects and the running of $\alpha_{\rm em}$.

\begin{figure}[p]
\vfill
\setlength{\unitlength}{1.0cm}
\begin{picture} (18.0,10.0)
\put (-2.0,0.0){\epsfig{figure=\figdir 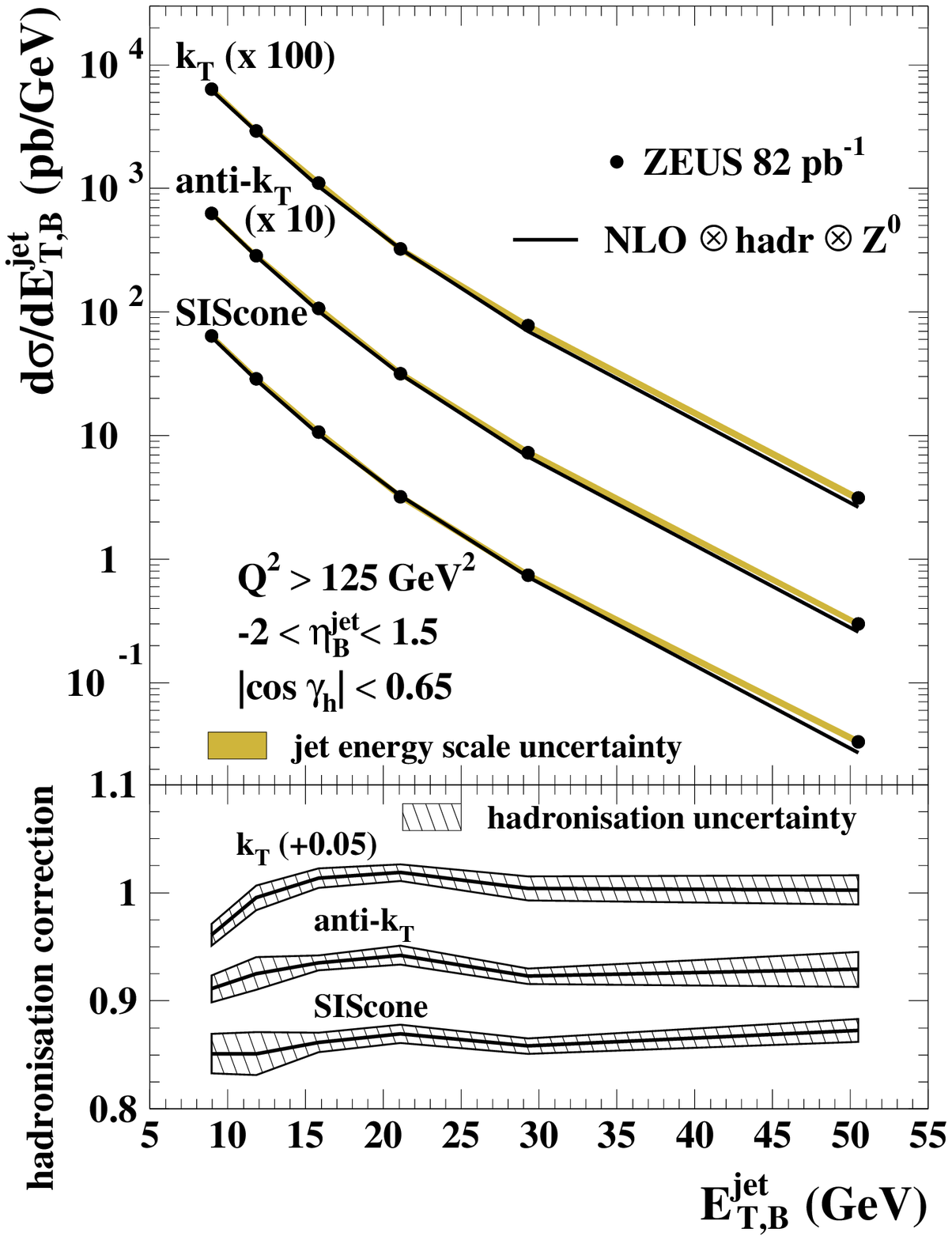,width=12cm}}
\put (7.0,0.0){\epsfig{figure=\figdir 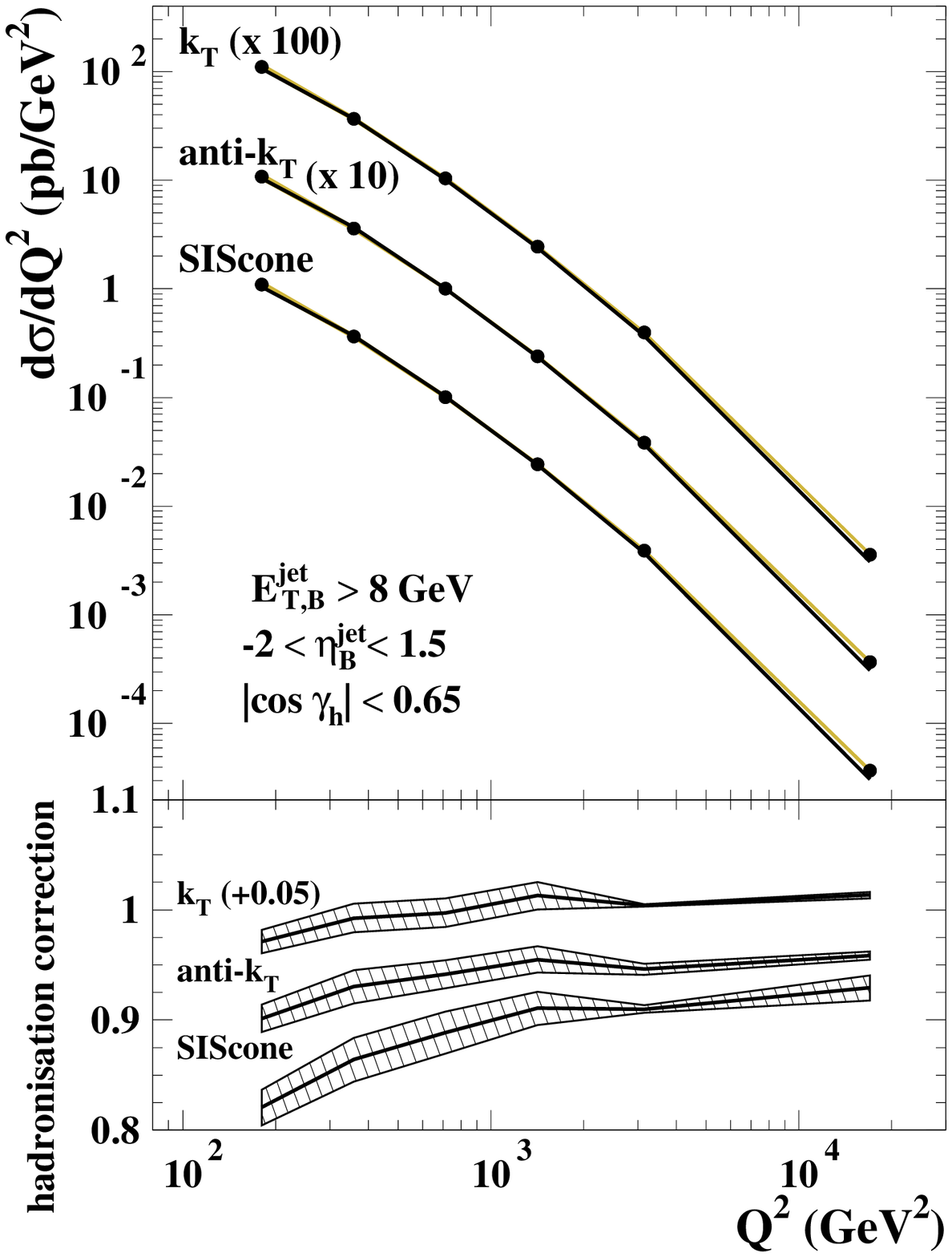,width=12cm}}
\put (0.0,0.0){\centerline{\epsfig{figure=\figdir 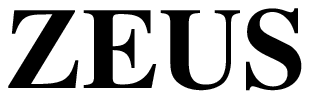,width=15cm}}}
\put (6.8,11.3){\bf\small (a)}
\put (15.8,11.3){\bf\small (b)}
\end{picture}
\caption
{\it 
The measured differential cross-sections (a) $\setjb$ and (b) $\sq2$
for inclusive-jet production (dots) using different jet
algorithms. The NLO QCD calculations with $\mu_R=\etjb$ (solid lines)
are also shown. The cross sections for the anti-$\kt$ and
$\kt$~\protect\cite{pl:b649:12} algorithms were multiplied by the scale
factors indicated in brackets to aid visibility. The error bars on the
data points are smaller than the marker size and are therefore not visible.
The shaded bands display the uncertainty due to the absolute energy
scale of the jets. The lower part of the figures shows the
hadronisation-correction factor applied to the NLO calculations
together with its uncertainty (hatched bands) for each jet algorithm;
the hadronisation-correction factor for the $\kt$ algorithm was
shifted by the value indicated in brackets to aid visibility.
}
\label{fig1}
\vfill
\end{figure}

The measurements of the inclusive-jet differential cross sections as
functions of $\etjb$ and $\q2$ are presented in Fig.~\ref{fig1} and
Tables~\ref{tabone} and \ref{tabtwo} for the anti-$\kt$ and SIScone
jet algorithms. For comparison, the measurements of inclusive-jet
cross sections with the $\kt$ algorithm~\cite{pl:b649:12} are also
shown. In Fig.~\ref{fig1}, each data point is plotted at the abscissa
at which the NLO QCD differential cross section was equal to its
bin-averaged value. The measured $\setjb$ ($\sq2$) exhibits a steep
fall-off over three (five) orders of magnitude for the jet algorithms
considered in the $\etjb$ ($\q2$) measured range. The measurements
using the three jet algorithms have a very similar shape and
normalisation.

\begin{figure}[p]
\vfill
\setlength{\unitlength}{1.0cm}
\begin{picture} (18.0,10.0)
\put (-2.0,0.0){\epsfig{figure=\figdir 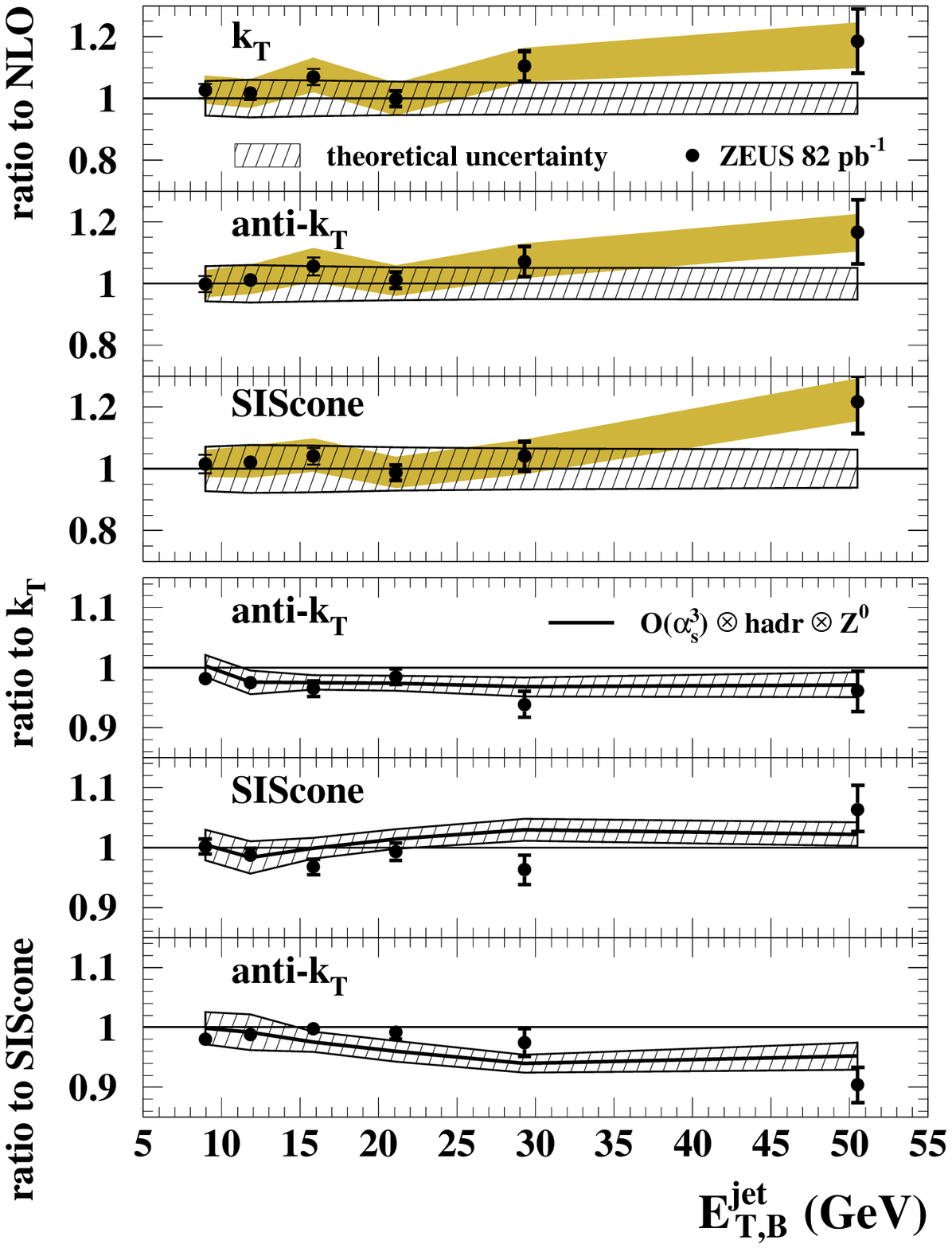,width=12cm}}
\put (7.0,0.0){\epsfig{figure=\figdir 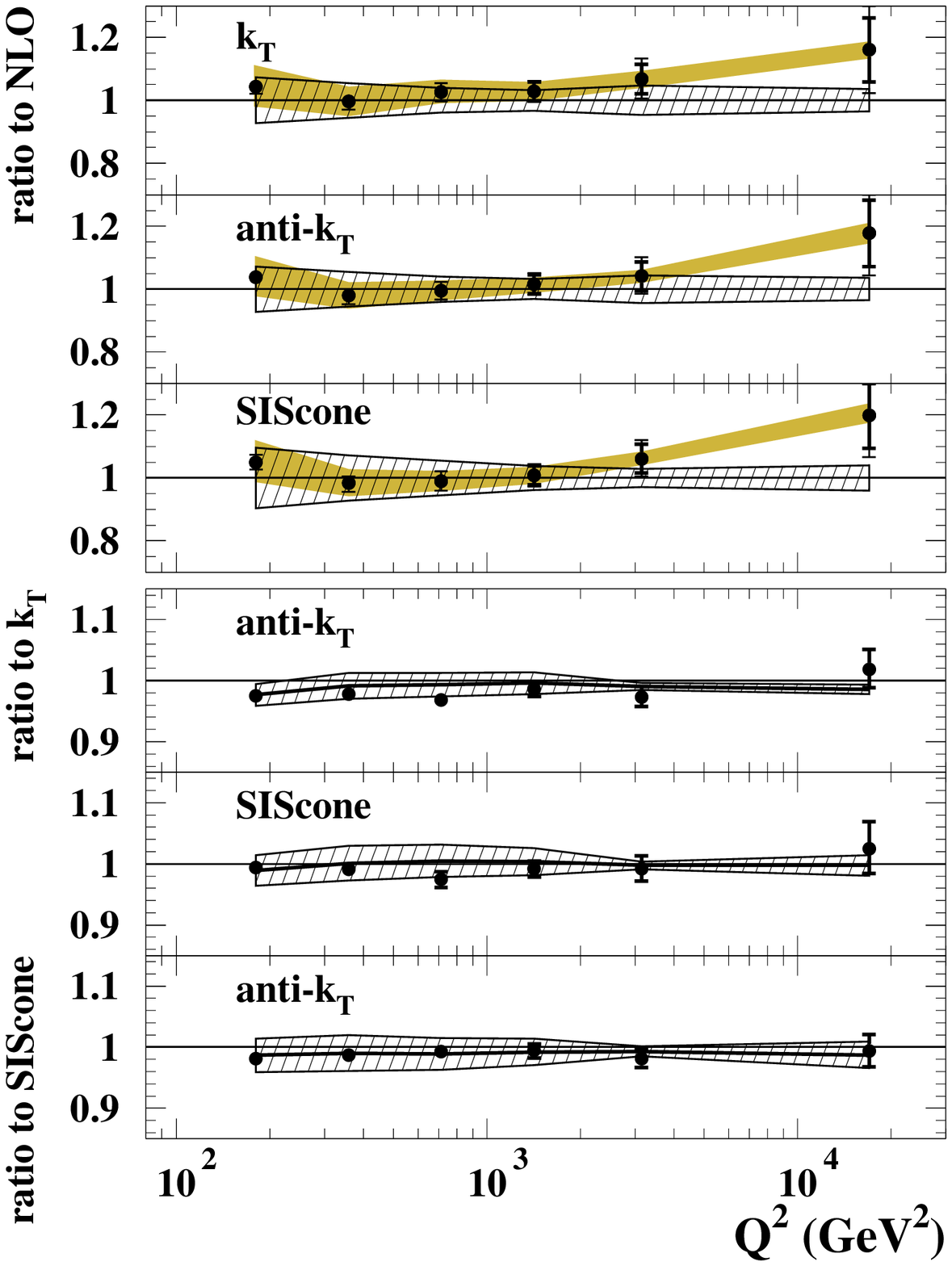,width=12cm}}
\put (0.0,-0.5){\centerline{\epsfig{figure=\figdir zeus.eps,width=15cm}}}
\put (0.5,11.0){\bf\small (a)}
\put (9.5,11.0){\bf\small (b)}
\put (0.5,5.7){\bf\small (c)}
\put (9.5,5.7){\bf\small (d)}
\end{picture}
\caption
{\it 
The ratios between the measured cross-sections (a) $\setjb$ and (b)
$\sq2$ and the NLO QCD calculations (dots). The inner error bars
represent the statistical uncertainty. The outer error bars show the
statistical and systematic uncertainties, not associated with the
uncertainty of the absolute energy scale of the jets, added in
quadrature. The hatched bands display the total theoretical
uncertainty and the shaded bands display the jet-energy scale
uncertainty. The ratios of the measured cross sections (dots)
anti-$\kt$/$\kt$, SIScone/$\kt$ and anti-$\kt$/SIScone as functions of
(c) $\etjb$ and (d) $\q2$. In these plots, the outer error bars
include also the uncertainty of the absolute energy scale of the jets.
The predicted ratios based on calculations which include up to
$\oasss$ terms are also shown (solid lines). The hatched bands display
the theoretical uncertainty on the ratio.
}
\label{fig2}
\vfill
\end{figure}

The NLO QCD predictions with $\mu_R=\etjb$ are compared to the
measurements in Fig.~\ref{fig1}. The hadronisation-correction factors
and their uncertainties are also shown. The ratios of the measured
differential cross sections to the NLO QCD calculations are shown in
Figs.~\ref{fig2}a and \ref{fig2}b separately for each jet algorithm. 
The measured differential cross sections are well reproduced by the
calculations, with similar precision in all cases. To study the
scale dependence, NLO QCD calculations using $\mu_R=Q$ were also
compared to the data (not shown); they also provide a good description
of the data. 

To compare in more detail the results of the different algorithms,
the ratios of the measured cross-sections anti-$\kt$/$\kt$,
SIScone/$\kt$ and anti-$\kt$/SIScone were investigated (see
Figs.~\ref{fig2}c and \ref{fig2}d). In the ratios, the statistical
correlations among the event samples as well as those among the jets
in the same event were taken into account in the estimation of the
statistical uncertainties. These ratios show that the cross sections
with the three jet algorithms are similar: the ratios as functions of
$\q2$ differ from unity by less than $3.2\%$; as functions of $\etjb$,
they differ by less than $3.6\%$ except at high $\etjb$, where the
difference is $10\%$. The pQCD predictions including up
to $\oasss$ terms for the ratios of cross sections are also 
shown in Figs.~\ref{fig2}c and \ref{fig2}d. In the estimation of the
total theoretical uncertainty of the predicted ratio,
all the theoretical contributions were assumed to be
correlated except that due to terms beyond $\oasss$ and that from the
modelling of the QCD cascade. The measured ratios are well described
by the calculations including up to $\oasss$ terms within the small
experimental and theoretical uncertainties, which are dominated by the
uncertainty on the modelling of the QCD cascade.

\begin{figure}[p]
\vfill
\setlength{\unitlength}{1.0cm}
\begin{picture} (18.0,10.0)
\put (-2.0,0.0){\epsfig{figure=\figdir 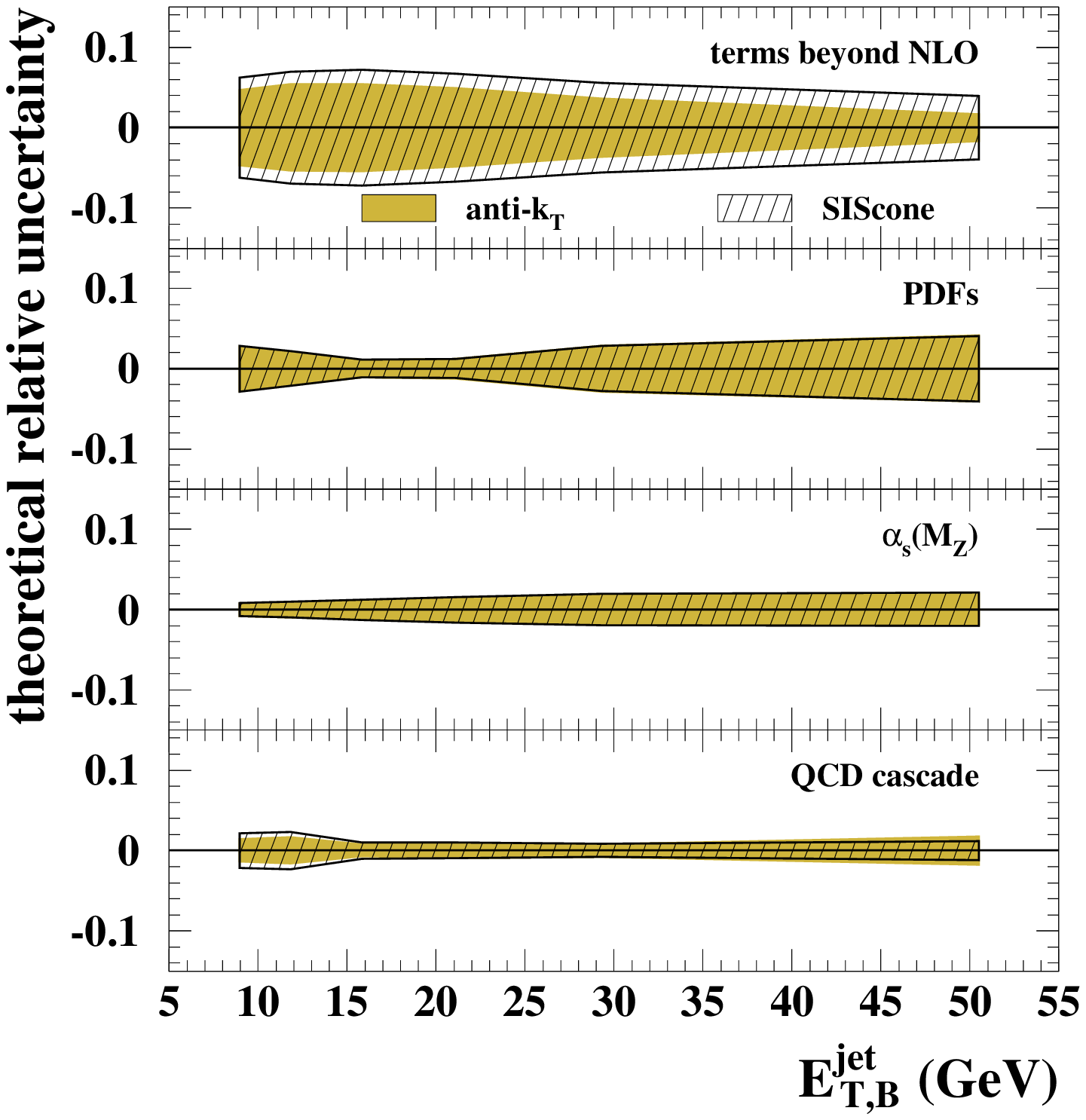,width=12cm}}
\put (7.0,0.0){\epsfig{figure=\figdir 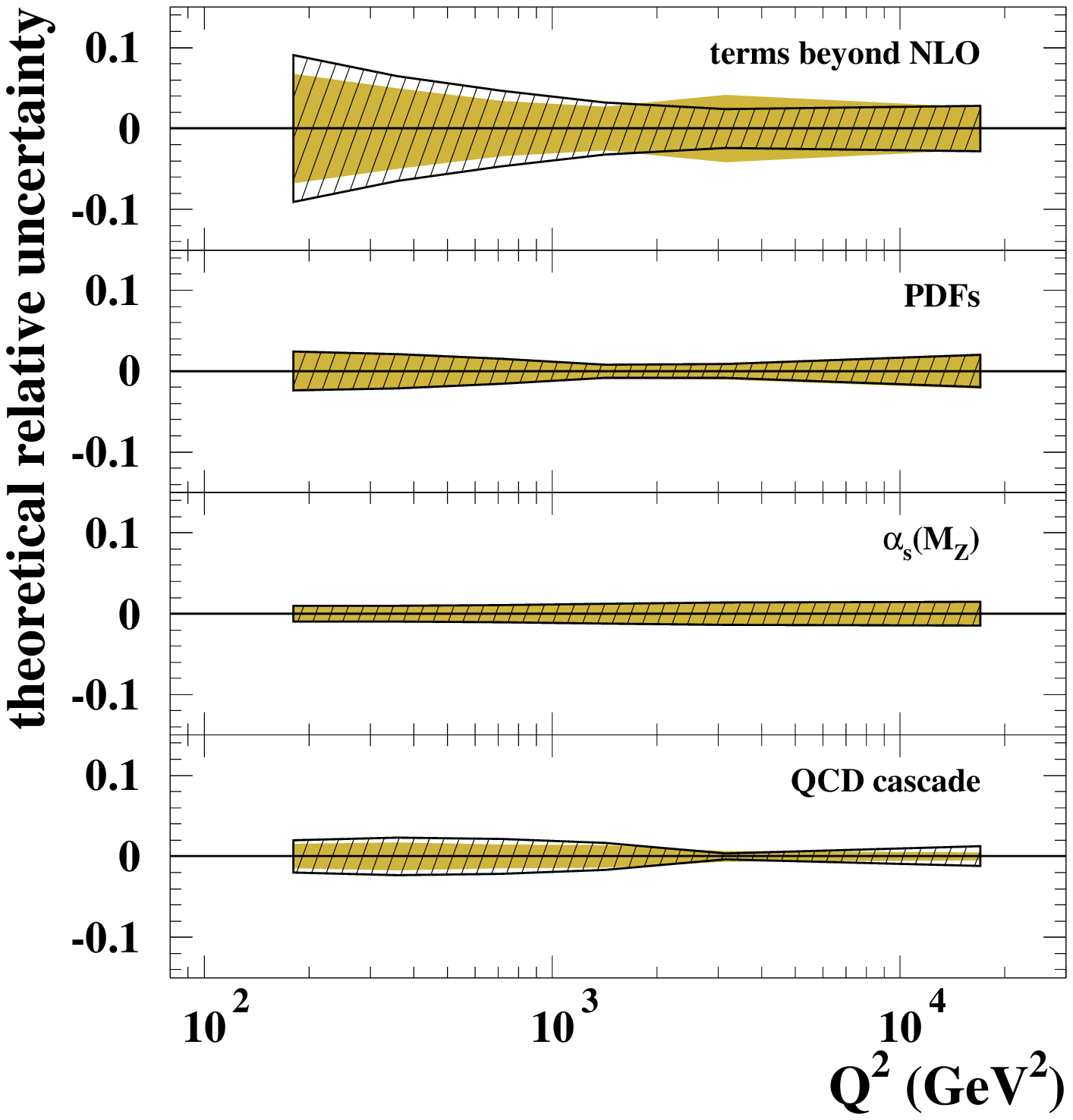,width=12cm}}
\put (0.0,-1.7){\centerline{\epsfig{figure=\figdir zeus.eps,width=15cm}}}
\put (0.6,9.8){\bf\small (a)}
\put (9.6,9.8){\bf\small (b)}
\end{picture}
\caption
{\it 
Overview of the theoretical relative uncertainties for the
inclusive-jet cross sections in the kinematic range of the measurements
as functions of (a) $\etjb$ and (b) $\q2$ for the anti-$\kt$ (shaded
areas) and SIScone (hatched areas) jet algorithms. Shown are the
relative uncertainties induced by the terms beyond NLO, those on
the proton PDFs, that on the value of $\asz$ and that on the modelling
of the QCD cascade.
}
\label{fig3}
\vfill
\end{figure}

Figure~\ref{fig3} shows the contributions to the theoretical
uncertainty of the inclusive-jet cross sections from terms beyond
NLO, from the uncertainty in the PDFs, from that on the value of
$\asz$ and from that on the modelling of the QCD cascade, separately
for the anti-$\kt$ and SIScone algorithms. The uncertainty coming from
the terms beyond NLO is dominant at low $\etjb$ and low $\q2$ in all
cases and somewhat higher for the predictions based on SIScone than
those for the anti-$\kt$ and $\kt$~\cite{pl:b649:12} algorithms. At
high $\etjb$, the PDF uncertainty is dominant in the case of the
anti-$\kt$ algorithm and of the same order as that arising from terms
beyond NLO for the SIScone algorithm. At high $\q2$, the uncertainty
due to terms beyond NLO is of the same order for both algorithms and
remains dominant. The uncertainty on the modelling of the QCD cascade
is somewhat higher for the SIScone algorithm than for anti-$\kt$, but
significantly smaller than that from terms beyond NLO. The total
theoretical uncertainty for the predictions using $\mu_R=Q$ is slightly
larger than that using $\mu_R=\etjb$.

\begin{figure}[p]
\vfill
\setlength{\unitlength}{1.0cm}
\begin{picture} (18.0,10.0)
\put (-2.0,0.0){\epsfig{figure=\figdir 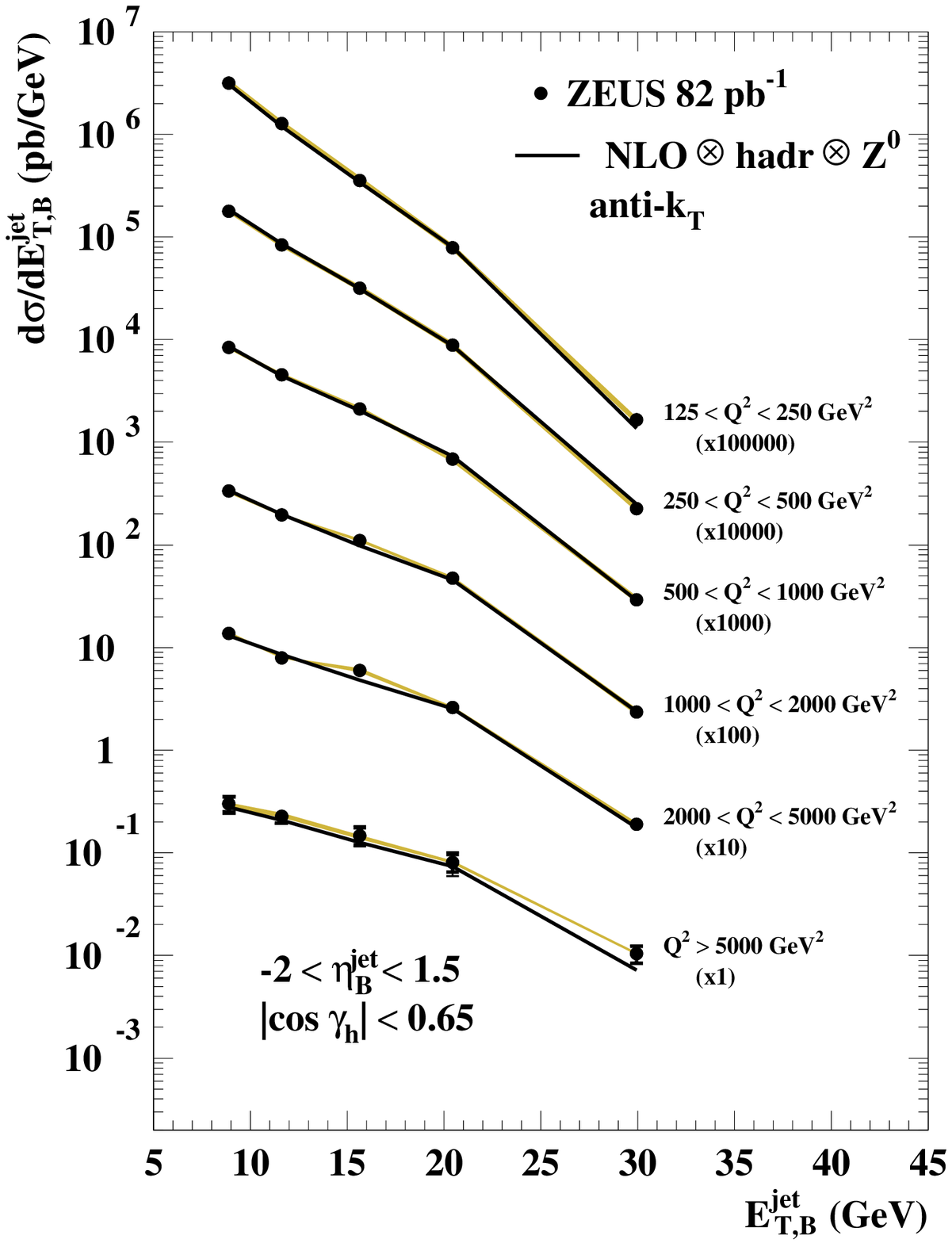,width=12cm}}
\put (7.0,0.0){\epsfig{figure=\figdir 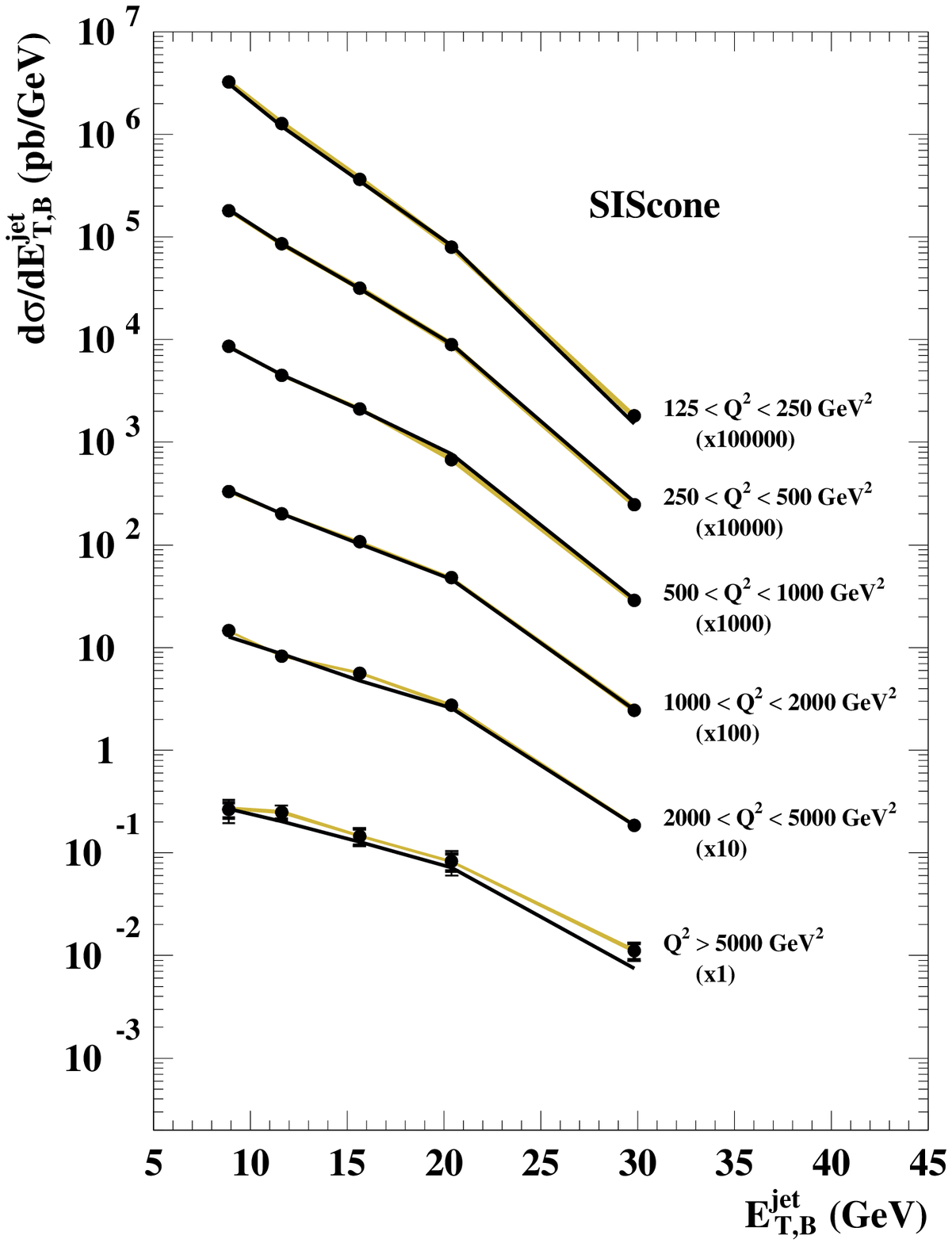,width=12cm}}
\put (0.0,-0.5){\centerline{\epsfig{figure=\figdir zeus.eps,width=15cm}}}
\put (6.7,9.0){\bf\small (a)}
\put (15.7,9.0){\bf\small (b)}
\end{picture}
\caption
{\it 
The measured differential cross-section $\setjb$ for inclusive-jet
production in different regions of $\q2$ (dots) using the (a)
anti-$\kt$ and (b) SIScone jet algorithms. The measured and predicted
cross sections have been multiplied by a scale factor as indicated in
brackets to aid visibility. Other details as in the captions to
Figs.~\ref{fig1} and \ref{fig2}.
}
\label{fig4}
\vfill
\end{figure}

\begin{figure}[p]
\vfill
\setlength{\unitlength}{1.0cm}
\begin{picture} (18.0,10.0)
\put (-2.0,1.0){\epsfig{figure=\figdir 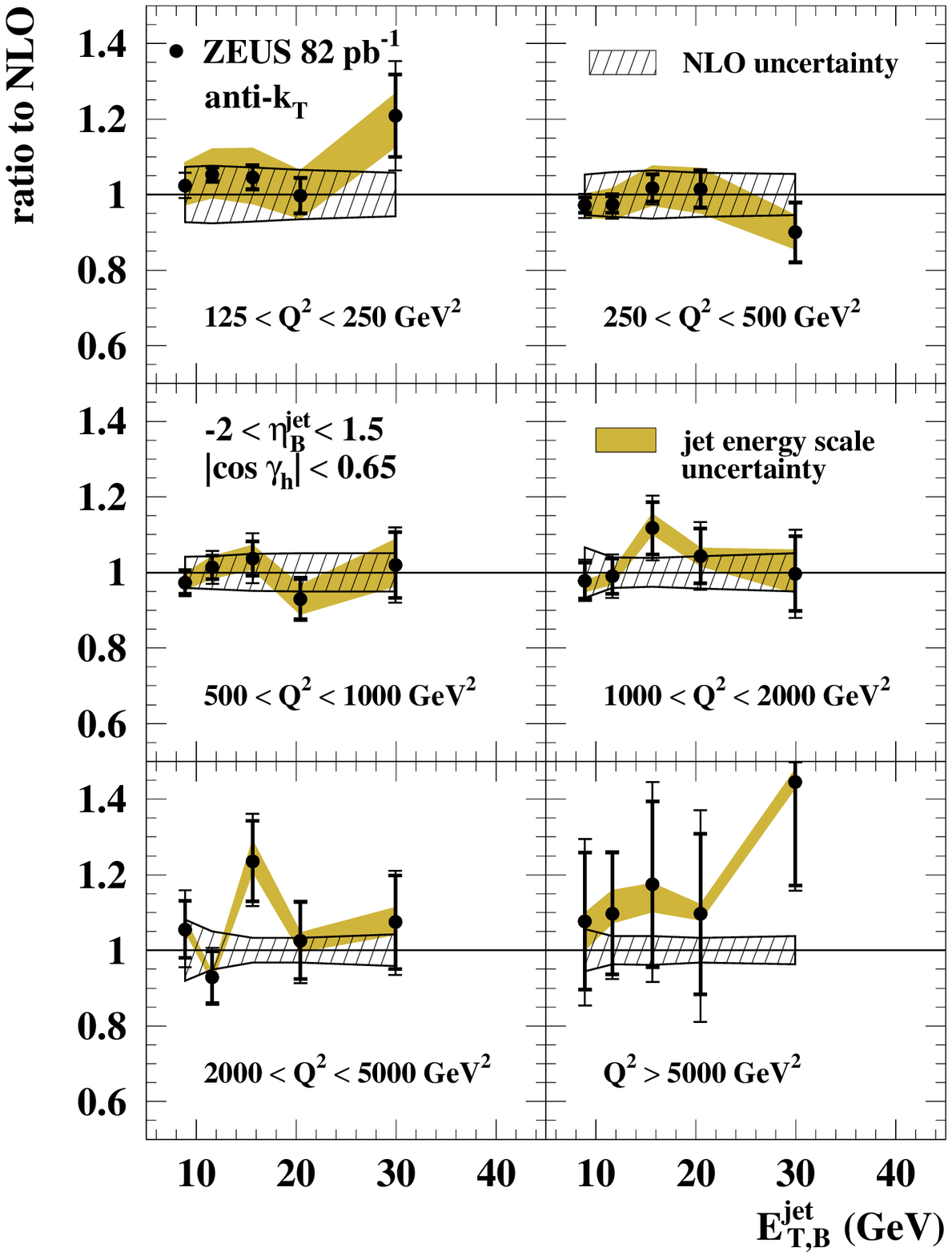,width=12cm}}
\put (7.0,1.0){\epsfig{figure=\figdir 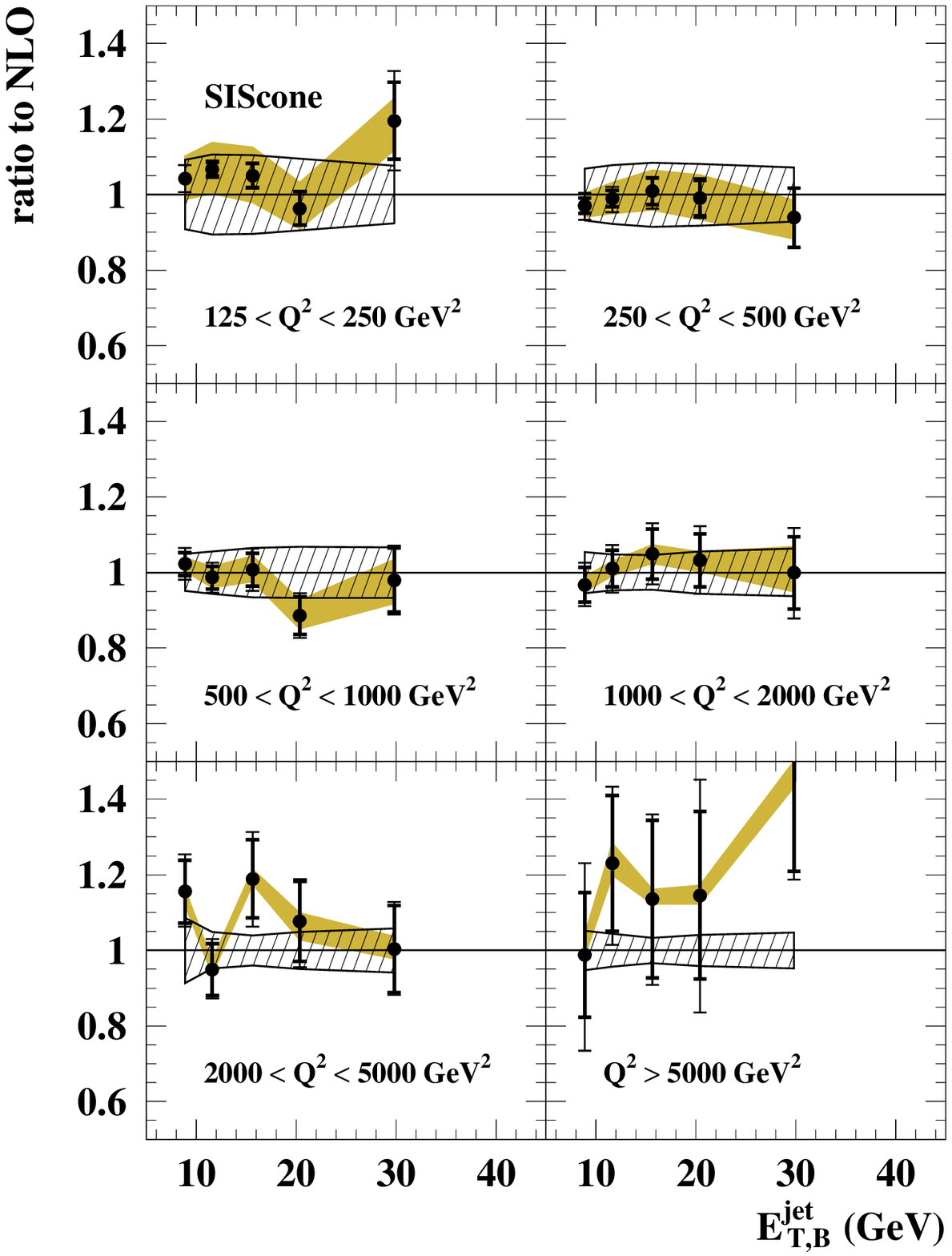,width=12cm}}
\put (0.0,0.5){\centerline{\epsfig{figure=\figdir zeus.eps,width=15cm}}}
\put (6.7,11.0){\bf\small (a)}
\put (15.7,11.0){\bf\small (b)}
\end{picture}
\vspace{-1.5cm}
\caption
{\it 
The ratios between the measured differential cross-sections
$\setjb$ presented in Fig.~\ref{fig4} and the NLO QCD calculations (dots). 
Other details as in the captions to Figs.~\ref{fig1} and \ref{fig2}.
}
\label{fig5}
\vfill
\end{figure}

The inclusive-jet cross sections for different regions of $\q2$ as a
function of $\etjb$ are presented in Fig.~\ref{fig4} and
Tables~\ref{tabthree} and \ref{tabfour} for both algorithms. 
The measured cross sections exhibit a steep fall-off within the $\etjb$ 
range considered. As $\q2$ increases, the $\etjb$ dependence of the 
cross section becomes less steep. The measurements have a similar
shape and normalisation for both jet algorithms and are similar to the
results obtained with the $\kt$ algorithm~\cite{np:b765:1}. The NLO QCD
predictions are compared to the measurements in Fig.~\ref{fig4}. The
data are well described by the predictions. Figure~\ref{fig5} shows
the ratios of the measured differential cross sections to the NLO QCD
calculations. The uncertainty of the NLO QCD calculations is also
shown: the uncertainties from the anti-$\kt$ predictions are of a
similar size as those encountered for the $\kt$
algorithm~\cite{np:b765:1}, and those for SIScone are somewhat
larger at low $\etjb$ and low $\q2$.

In summary, it is concluded that the data for the $\kt$, anti-$\kt$
and SIScone jet algorithms are well described by the NLO QCD
calculations with similar experimental and theoretical
precision. Furthermore, the measured ratios are well described by the
predictions including up to $\oasss$ terms, demonstrating the ability
of the pQCD calculations with up to four partons in the final state
to account adequately for the differences between the jet algorithms.

\subsection{Determination of {\boldmath $\asz$}}
The measured differential cross-sections $\sq2$ for $\q2>500$~\g2\
were used to determine values of $\asz$ using the method 
presented previously~\cite{pl:b547:164}. The NLO QCD calculations were
performed using the program {\sc Disent} with five sets of
ZEUS-S proton PDFs which were determined from global fits assuming
different values of $\asz$, namely 
$\asz=0.115,\ 0.117,\ 0.119,\ 0.121$ and $0.123$. The value of $\asz$
used in each calculation was that associated with the corresponding
set of PDFs. The $\asz$ dependence of the predicted cross sections in
each bin $i$ of $\q2$ was parameterised according to 

$$\left [ \sq2(\asz) \right ]_i=C_1^i\asz+C_2^i\as^2(\mz),$$

where $C_1^i$ and $C_2^i$ were determined from a $\chi^2$ fit to
the NLO QCD calculations. The value of $\asz$ was determined by a
$\chi^2$ fit to the measured $\sq2$ values.

The uncertainties on the extracted values of $\asz$ due to the
experimental systematic uncertainties were evaluated by repeating the
analysis for each systematic check presented in Section~\ref{expunc}.
The overall normalisation uncertainty from the luminosity determination
was also considered. The largest contribution to the experimental
uncertainty comes from the jet energy scale and amounts to $\pm 1.9\%$
on $\asz$ for both algorithms ($\pm 2\%$ for
$\kt$~\cite{pl:b649:12}). The theoretical uncertainties were evaluated
as described in Section~\ref{thun}. The largest contribution was the
theoretical uncertainty arising from terms beyond NLO, which was
estimated by using the method proposed by Jones 
\etal~\cite{jhep:0312:007}, and amounted to 
$\pm 1.5\%$ for both algorithms\footnote{The theoretical uncertainty
  arising from terms beyond NLO estimated by refitting the data using
  calculations based on $\mu_R=2\etjb$ or $\etjb/2$ amounts to 
  $_{-0.0}^{+2.1}\ (_{-0.7}^{+3.1})\%$ for the anti-$\kt$ (SIScone)
  algorithm.}
($\pm 1.5\%$ for $\kt$~\cite{pl:b649:12}).
The uncertainty due to the proton PDFs was $\pm 0.7\ (\pm 0.8)\%$ for the
anti-$\kt$ (SIScone) algorithm ($\pm 0.7\%$ for 
$\kt$~\cite{pl:b649:12}). The uncertainty arising from the hadronisation 
effects amounted to $\pm 0.9\ (\pm 1.2)\%$ for the anti-$\kt$ (SIScone) 
algorithm ($\pm 0.8\%$ for $\kt$~\cite{pl:b649:12}). Thus, the
performance of the three jet algorithms is very similar.

As a cross-check, $\asz$ was determined by using NLO QCD calculations
based on the CTEQ6.1~\cite{jhep:0207:012,*jhep:0310:046}
(MSTW2008~\cite{epj:c63:189}) sets of proton PDFs: the values
obtained are consistent within $\pm 1.2\ (\pm 1.0)\%$ with those based on
ZEUS-S. The uncertainty arising from the proton PDFs was estimated to
be $\pm 1.5\ (\pm 0.7,\ \pm 0.4)\%$ using the results of the CTEQ6.1
(MSTW2008nlo90cl, MSTW2008nlo68cl) analysis.

The values of $\asz$ obtained from the measured $\sq2$ for
$\q2>500$~\g2\ are

\begin{eqnarray}
\asz|_{{\rm anti-}\kt} &=& 0.1188\pm 0.0014\ {\rm (stat.)}\ ^{+0.0033}_{-0.0032}\ {\rm (exp.)}\ ^{+0.0022}_{-0.0022}\ {\rm (th.)}\ 
{\rm and}\nonumber\\
\asz|_{\rm SIScone} &=& 0.1186\pm 0.0013\ {\rm (stat.)}\ ^{+0.0034}_{-0.0032}\ {\rm (exp.)}\ ^{+0.0025}_{-0.0025}\ {\rm (th.)}.\nonumber
\end{eqnarray}

These values of $\asz$ are consistent with that obtained using the
$\kt$ algorithm~\cite{pl:b649:12}, 
\begin{center}
$\asmzn{0.1207}{0.0014}{0.0033}{0.0035}{0.0023}{0.0022}{\kt}$,
\end{center}
with the current world average of $0.1189\pm 0.0010$~\cite{jp:g26:r27},
with the results obtained in $\pp$ 
collisions~\cite{prl:88:042001,*pr:d80:111107}
as well as with the HERA average of 
$0.1186\pm 0.0051$~\cite{proc:dis:2005:689}. It should be noted that the
differences between the central values of $\asz$ obtained using the
three jet algorithms are comparable to the uncertainties due to
higher-order terms in the calculations. It is observed that the
precision in $\asz$ obtained with the $\kt$, anti-$\kt$ and SIScone
jet algorithms is very similar, and comparable to those obtained in
$\ele$ interactions~\cite{jp:g26:r27}.

\section{Summary and conclusions}

For the first time, differential cross sections for
inclusive-jet production in neutral current deep inelastic $ep$
scattering were measured using the anti-$\kt$ and SIScone jet
algorithms with $R=1$. The cross sections correspond to a centre-of-mass
energy of 318~GeV and refer to jets of hadrons with $\etjb>8$~GeV and
$-2<\etajb<1.5$ identified in the Breit frame with the anti-$\kt$ or
SIScone jet algorithms. The measurements are given in the kinematic
region of $\q2>125$~\g2\ and $|\cgh|<0.65$.

A detailed comparison between the measurements as functions of
$\etjb$ and $\q2$ for both algorithms and those from a previous
analysis based on the $\kt$ jet reconstruction was performed. The
measured cross sections for the three jet algorithms have similar shapes
and normalisations.

The NLO QCD calculations of inclusive-jet cross sections and their
uncertainties for the different algorithms were also compared:
the data are well described by the predictions; the calculations for the
anti-$\kt$ algorithm have a similar precision as that of the $\kt$
whereas those for the SIScone are somewhat less precise due to the
contribution from terms beyond NLO.

Measurements of the ratios of cross sections using different
jet algorithms were also presented and compared to calculations
including up to $\oasss$ terms. The measured ratios are well
reproduced by the predictions, demonstrating the ability of the pQCD 
calculations including up to four partons in the final state
to account adequately for the differences between the jet algorithms.

Values of $\asz$ were extracted from the measured
inclusive-jet differential cross sections using each algorithm.
QCD fits of the cross-sections $\sq2$ for $\q2>500$~\g2\ yield the
following values of $\asz$:

\begin{eqnarray}
\asz|_{{\rm anti-}\kt} &=& 0.1188\pm 0.0014\ {\rm (stat.)}\ ^{+0.0033}_{-0.0032}\ {\rm (exp.)}\ ^{+0.0022}_{-0.0022}\ {\rm (th.)}\ 
{\rm and}\nonumber\\
\asz|_{\rm SIScone} &=& 0.1186\pm 0.0013\ {\rm (stat.)}\ ^{+0.0034}_{-0.0032}\ {\rm (exp.)}\ ^{+0.0025}_{-0.0025}\ {\rm (th.)}.\nonumber
\end{eqnarray}
These values are consistent with each other and with that obtained
from the $\kt$ analysis with a similar precision.

\newpage
\vspace{0.5cm}
\noindent {\Large\bf Acknowledgements}
\vspace{0.3cm}

We thank the DESY Directorate for their strong support and
encouragement. The remarkable achievements of the HERA machine group
were essential for the successful completion of this work and are
greatly appreciated. We are grateful for the support of the DESY
computing and network services. The design, construction and
installation of the ZEUS detector have been made possible owing to the
ingenuity and effort of many people who are not listed as authors. We
would like to thank M. Cacciari, Z. Nagy, G. P. Salam and G. Soyez 
for useful discussions.

\providecommand{\etal}{et al.\xspace}
\providecommand{\coll}{Collaboration}
\catcode`\@=11
\def\@bibitem#1{%
\ifmc@bstsupport
  \mc@iftail{#1}%
    {;\newline\ignorespaces}%
    {\ifmc@first\else.\fi\orig@bibitem{#1}}
  \mc@firstfalse
\else
  \mc@iftail{#1}%
    {\ignorespaces}%
    {\orig@bibitem{#1}}%
\fi}%
\catcode`\@=12

\clearpage
\newpage
\begin{table}
\begin{center}
    \begin{tabular}{|c||cccc||c||c|}
\hline
  $\etjb$ bin
& $d\sigma/d\etjb$
& $\delta_{\rm stat}$
& $\delta_{\rm syst}$
& $\delta_{\rm ES}$
& $C_{\rm QED}$
& $C_{\rm had}\cdot C_{\z0}$\\
  (GeV)
& (pb/GeV)
&
&
&
&
& \\
\hline
\multicolumn{7}{c}{anti-$\kt$} \\
\hline
$8-10  $ & $62.84  $ & $\pm 0.69  $ & $_{-1.49}^{+1.40}    $ & $_{-2.46}^{+2.67}    $ & $0.95$ & $0.93$ \\
$10-14 $ & $28.56  $ & $\pm 0.34  $ & $_{-0.48}^{+0.39}    $ & $_{-1.25}^{+1.31}    $ & $0.97$ & $0.93$ \\
$14-18 $ & $10.69  $ & $\pm 0.20  $ & $_{-0.23}^{+0.20}    $ & $_{-0.49}^{+0.57}    $ & $0.96$ & $0.92$ \\
$18-25 $ & $ 3.186 $ & $\pm 0.082 $ & $_{-0.040}^{+0.041}  $ & $_{-0.154}^{+0.143}  $ & $0.95$ & $0.92$ \\
$25-35 $ & $ 0.725 $ & $\pm 0.032 $ & $_{-0.014}^{+0.014}  $ & $_{-0.034}^{+0.038}  $ & $0.94$ & $0.94$ \\
$35-100$ & $ 0.0300$ & $\pm 0.0026$ & $_{-0.0006}^{+0.0005}$ & $_{-0.0015}^{+0.0014}$ & $1.06$ & $0.92$ \\
\hline
\multicolumn{7}{c}{SIScone} \\
\hline
$8-10  $ & $64.08$ & $\pm 0.69$ & $_{-1.82}^{+1.74}$ & $_{-2.54}^{+2.73}$ & $0.95$ & $0.87$ \\
$10-14 $ & $28.92$ & $\pm 0.34$ & $_{-0.49}^{+0.42}$ & $_{-1.29}^{+1.42}$ & $0.96$ & $0.85$ \\
$14-18 $ & $10.72$ & $\pm 0.20$ & $_{-0.22}^{+0.16}$ & $_{-0.51}^{+0.55}$ & $0.96$ & $0.85$ \\
$18-25 $ & $ 3.211$ & $\pm 0.081$ & $_{-0.043}^{+0.045}$ & $_{-0.149}^{+0.161}$ & $0.95$ & $0.85$ \\
$25-35 $ & $ 0.744$ & $\pm 0.033$ & $_{-0.017}^{+0.017}$ & $_{-0.039}^{+0.036}$ & $0.93$ & $0.88$ \\
$35-100$ & $ 0.0332$ & $\pm 0.0028$ & $_{-0.0008}^{+0.0008}$ & $_{-0.0016}^{+0.0020}$ & $1.06$ & $0.86$ \\
\hline
    \end{tabular}
 \caption{
The measured differential cross-sections $\setjb$ for inclusive-jet
production using different jet algorithms. The statistical,
uncorrelated systematic and jet-energy-scale ({\rm ES}) uncertainties
are shown separately. The multiplicative corrections applied to the
data to correct for QED radiative effects, $C_{\rm QED}$, and the
corrections for hadronisation and $\z0$ effects to be applied to the
parton-level NLO QCD calculations, $C_{\rm had}\cdot C_{\z0}$, are
shown in the last two columns.}
 \label{tabone}
\end{center}
\end{table}

\clearpage
\newpage
\begin{table}
\begin{center}
    \begin{tabular}{|c||cccc||c||c|}
\hline
  $\q2$ bin
& $d\sigma/d\q2$
& $\delta_{\rm stat}$
& $\delta_{\rm syst}$
& $\delta_{\rm ES}$
& $C_{\rm QED}$
& $C_{\rm had}\cdot C_{\z0}$\\
  (\g2)
& (pb/\g2)
&
&
&
&
& \\
\hline
\multicolumn{7}{c}{anti-$\kt$} \\
\hline
$125-250    $ & $1.078    $ & $\pm 0.012    $ & $_{-0.018}^{+0.017}        $ & $_{-0.060}^{+0.067}        $ & $0.97$ & $0.90$ \\
$250-500    $ & $0.3589   $ & $\pm 0.0053   $ & $_{-0.0084}^{+0.0055}      $ & $_{-0.0142}^{+0.0147}      $ & $0.95$ & $0.94$ \\
$500-1000   $ & $0.1004   $ & $\pm 0.0020   $ & $_{-0.0019}^{+0.0019}      $ & $_{-0.0028}^{+0.0030}      $ & $0.95$ & $0.94$ \\
$1000-2000  $ & $0.02406  $ & $\pm 0.00072  $ & $_{-0.00033}^{+0.00044}    $ & $_{-0.00058}^{+0.00045}    $ & $0.94$ & $0.96$ \\
$2000-5000  $ & $0.00385  $ & $\pm 0.00017  $ & $_{-0.00012}^{+0.00015}    $ & $_{-0.00007}^{+0.00007}    $ & $0.94$ & $0.97$ \\
$5000-100000$ & $36.6\cdot 10^{-6}$ & $\pm 3.3\cdot 10^{-6}$ & $_{-2.6}^{+2.2}\cdot 10^{-6}$ & $_{-0.9}^{+1.0}\cdot 10^{-6}$ & $0.98$ & $0.95$ \\
\hline
\multicolumn{7}{c}{SIScone} \\
\hline
$125-250    $ & $1.099    $ & $\pm 0.012    $ & $_{-0.021}^{+0.020}        $ & $_{-0.063}^{+0.069}        $ & $0.97$ & $0.82$ \\
$250-500    $ & $0.3637   $ & $\pm 0.0053   $ & $_{-0.0089}^{+0.0062}      $ & $_{-0.0143}^{+0.0156}      $ & $0.95$ & $0.87$ \\
$500-1000   $ & $0.1011   $ & $\pm 0.0020   $ & $_{-0.0024}^{+0.0025}      $ & $_{-0.0028}^{+0.0030}      $ & $0.95$ & $0.89$ \\
$1000-2000  $ & $0.02419  $ & $\pm 0.00071  $ & $_{-0.00045}^{+0.00048}    $ & $_{-0.00057}^{+0.00060}    $ & $0.94$ & $0.92$ \\
$2000-5000  $ & $0.00393  $ & $\pm 0.00017  $ & $_{-0.00012}^{+0.00014}    $ & $_{-0.00007}^{+0.00007}    $ & $0.94$ & $0.93$ \\
$5000-100000$ & $36.8\cdot 10^{-6}$ & $\pm 3.2\cdot 10^{-6}$ & $_{-2.5}^{+1.8}\cdot 10^{-6}$ & $_{-0.6}^{+1.1}\cdot 10^{-6}$ & $0.97$ & $0.92$ \\
\hline
    \end{tabular}
 \caption{
   The measured differential cross-sections $\sq2$ for inclusive-jet
   production using different jet algorithms. Other details as in
   the caption to Table~\ref{tabone}.}
 \label{tabtwo}
\end{center}
\end{table}

\clearpage
\newpage
\begin{table}
\vspace*{-3.cm}
\begin{center}
    \begin{tabular}{|c||cccc||c||c|}
\hline
  $\etjb$ bin
& $d\sigma/d\etjb$
& $\delta_{\rm stat}$
& $\delta_{\rm syst}$
& $\delta_{\rm ES}$
& $C_{\rm QED}$
& $C_{\rm had}\cdot C_{\z0}$\\
  (GeV)
& (pb/GeV)
&
&
&
&
& \\
\hline
\multicolumn{7}{c}{$125<\q2<250$~\g2} \\
\hline
$8-10  $ & $31.60 $ & $\pm 0.49  $ & $_{-0.93}^{+0.90}    $ & $_{-1.56}^{+1.87}    $ & $0.95$ & $0.91$ \\
$10-14 $ & $12.67 $ & $\pm 0.22  $ & $_{-0.15}^{+0.13}    $ & $_{-0.74}^{+0.80}    $ & $0.99$ & $0.90$ \\
$14-18 $ & $3.56  $ & $\pm 0.11  $ & $_{-0.05}^{+0.03}    $ & $_{-0.24}^{+0.26}    $ & $0.96$ & $0.88$ \\
$18-25 $ & $0.790 $ & $\pm 0.037 $ & $_{-0.012}^{+0.013}  $ & $_{-0.047}^{+0.051}  $ & $0.95$ & $0.86$ \\
$25-100$ & $0.0165$ & $\pm 0.0015$ & $_{-0.0013}^{+0.0013}$ & $_{-0.0011}^{+0.0008}$ & $1.02$ & $0.85$ \\
\hline
\multicolumn{7}{c}{$250<\q2<500$~\g2} \\
\hline
$8-10  $ & $17.90 $ & $\pm 0.38  $ & $_{-0.50}^{+0.41}    $ & $_{-0.58}^{+0.52}    $ & $0.94$ & $0.94$ \\
$10-14 $ & $8.37  $ & $\pm 0.19  $ & $_{-0.26}^{+0.18}    $ & $_{-0.32}^{+0.36}    $ & $0.96$ & $0.94$ \\
$14-18 $ & $3.17  $ & $\pm 0.11  $ & $_{-0.08}^{+0.04}    $ & $_{-0.14}^{+0.18}    $ & $0.97$ & $0.94$ \\
$18-25 $ & $0.882 $ & $\pm 0.042 $ & $_{-0.015}^{+0.016}  $ & $_{-0.055}^{+0.047}  $ & $0.92$ & $0.89$ \\
$25-100$ & $0.0225$ & $\pm 0.0020$ & $_{-0.0006}^{+0.0006}$ & $_{-0.0011}^{+0.0011}$ & $0.94$ & $0.95$ \\
\hline
\multicolumn{7}{c}{$500<\q2<1000$~\g2} \\
\hline
$8-10  $ & $8.36  $ & $\pm 0.26  $ & $_{-0.18}^{+0.17}    $ & $_{-0.14}^{+0.19}    $ & $0.96$ & $0.95$ \\
$10-14 $ & $4.52  $ & $\pm 0.14  $ & $_{-0.13}^{+0.14}    $ & $_{-0.13}^{+0.12}    $ & $0.94$ & $0.92$ \\
$14-18 $ & $2.116 $ & $\pm 0.092 $ & $_{-0.097}^{+0.099}  $ & $_{-0.058}^{+0.069}  $ & $0.93$ & $0.95$ \\
$18-25 $ & $0.682 $ & $\pm 0.039 $ & $_{-0.016}^{+0.017}  $ & $_{-0.029}^{+0.026}  $ & $0.97$ & $0.95$ \\
$25-100$ & $0.0293$ & $\pm 0.0025$ & $_{-0.0014}^{+0.0014}$ & $_{-0.0016}^{+0.0019}$ & $0.94$ & $0.98$ \\
\hline
\multicolumn{7}{c}{$1000<\q2<2000$~\g2} \\
\hline
$8-10  $ & $3.34  $ & $\pm 0.16  $ & $_{-0.08}^{+0.10}    $ & $_{-0.10}^{+0.00}    $ & $0.93$ & $0.97$ \\
$10-14 $ & $1.967 $ & $\pm 0.093 $ & $_{-0.065}^{+0.070}  $ & $_{-0.039}^{+0.025}  $ & $0.93$ & $0.96$ \\
$14-18 $ & $1.099 $ & $\pm 0.068 $ & $_{-0.048}^{+0.049}  $ & $_{-0.015}^{+0.036}  $ & $0.95$ & $0.94$ \\
$18-25 $ & $0.477 $ & $\pm 0.033 $ & $_{-0.024}^{+0.024}  $ & $_{-0.011}^{+0.009}  $ & $0.94$ & $1.00$ \\
$25-100$ & $0.0237$ & $\pm 0.0023$ & $_{-0.0015}^{+0.0015}$ & $_{-0.0012}^{+0.0015}$ & $0.99$ & $0.99$ \\
\hline
\multicolumn{7}{c}{$2000<\q2<5000$~\g2} \\
\hline
$8-10  $ & $1.374 $ & $\pm 0.099 $ & $_{-0.084}^{+0.094}  $ & $_{-0.004}^{+0.034}  $ & $0.91$ & $0.89$ \\
$10-14 $ & $0.797 $ & $\pm 0.059 $ & $_{-0.022}^{+0.030}  $ & $_{-0.009}^{+0.009}  $ & $0.93$ & $0.96$ \\
$14-18 $ & $0.598 $ & $\pm 0.051 $ & $_{-0.026}^{+0.032}  $ & $_{-0.012}^{+0.025}  $ & $1.01$ & $0.99$ \\
$18-25 $ & $0.261 $ & $\pm 0.026 $ & $_{-0.012}^{+0.006}  $ & $_{-0.007}^{+0.005}  $ & $0.97$ & $1.05$ \\
$25-100$ & $0.0189$ & $\pm 0.0022$ & $_{-0.0011}^{+0.0010}$ & $_{-0.0006}^{+0.0007}$ & $0.90$ & $1.02$ \\
\hline
\multicolumn{7}{c}{$5000<\q2<100000$~\g2} \\
\hline
$8-10  $ & $0.301 $ & $\pm 0.051 $ & $_{-0.037}^{+0.033}  $ & $_{-0.021}^{+0.005}  $ & $0.94$ & $0.89$ \\
$10-14 $ & $0.229 $ & $\pm 0.034 $ & $_{-0.013}^{+0.007}  $ & $_{-0.005}^{+0.013}  $ & $0.96$ & $0.93$ \\
$14-18 $ & $0.148 $ & $\pm 0.028 $ & $_{-0.017}^{+0.020}  $ & $_{-0.009}^{+0.000}  $ & $1.01$ & $0.92$ \\
$18-25 $ & $0.081 $ & $\pm 0.016 $ & $_{-0.014}^{+0.013}  $ & $_{-0.001}^{+0.002}  $ & $0.98$ & $0.99$ \\
$25-100$ & $0.0104$ & $\pm 0.0020$ & $_{-0.0007}^{+0.0007}$ & $_{-0.0001}^{+0.0002}$ & $1.00$ & $1.04$ \\
\hline
    \end{tabular}
 \caption{
   The measured differential cross-section $\setjb$ for inclusive-jet
   production in different regions of $\q2$ using the anti-$\kt$ jet
   algorithm. Other details as in the caption to Table~\ref{tabone}.}
 \label{tabthree}
\end{center}
\end{table}

\clearpage
\newpage
\begin{table}
\vspace*{-3.cm}
\begin{center}
    \begin{tabular}{|c||cccc||c||c|}
\hline
  $\etjb$ bin
& $d\sigma/d\etjb$
& $\delta_{\rm stat}$
& $\delta_{\rm syst}$
& $\delta_{\rm ES}$
& $C_{\rm QED}$
& $C_{\rm had}\cdot C_{\z0}$\\
  (GeV)
& (pb/GeV)
&
&
&
&
& \\
\hline
\multicolumn{7}{c}{$125<\q2<250$~\g2} \\
\hline
$8-10  $ & $32.36 $ & $\pm 0.49  $ & $_{-1.02}^{+0.99}    $ & $_{-1.67}^{+1.86}    $ & $0.96$ & $0.84$ \\
$10-14 $ & $12.79 $ & $\pm 0.22  $ & $_{-0.18}^{+0.17}    $ & $_{-0.78}^{+0.84}    $ & $0.98$ & $0.81$ \\
$14-18 $ & $3.66  $ & $\pm 0.11  $ & $_{-0.05}^{+0.04}    $ & $_{-0.24}^{+0.26}    $ & $0.97$ & $0.80$ \\
$18-25 $ & $0.792 $ & $\pm 0.037 $ & $_{-0.008}^{+0.009}  $ & $_{-0.044}^{+0.056}  $ & $0.95$ & $0.79$ \\
$25-100$ & $0.0180$ & $\pm 0.0015$ & $_{-0.0013}^{+0.0013}$ & $_{-0.0012}^{+0.0009}$ & $1.00$ & $0.80$ \\
\hline
\multicolumn{7}{c}{$250<\q2<500$~\g2} \\
\hline
$8-10  $ & $18.02 $ & $\pm 0.37  $ & $_{-0.60}^{+0.53}    $ & $_{-0.55}^{+0.59}    $ & $0.94$ & $0.89$ \\
$10-14 $ & $8.55  $ & $\pm 0.19  $ & $_{-0.24}^{+0.19}    $ & $_{-0.33}^{+0.37}    $ & $0.96$ & $0.87$ \\
$14-18 $ & $3.16  $ & $\pm 0.11  $ & $_{-0.09}^{+0.04}    $ & $_{-0.15}^{+0.17}    $ & $0.97$ & $0.86$ \\
$18-25 $ & $0.893 $ & $\pm 0.042 $ & $_{-0.021}^{+0.021}  $ & $_{-0.050}^{+0.055}  $ & $0.92$ & $0.82$ \\
$25-100$ & $0.0247$ & $\pm 0.0021$ & $_{-0.0004}^{+0.0004}$ & $_{-0.0014}^{+0.0012}$ & $0.94$ & $0.87$ \\
\hline
\multicolumn{7}{c}{$500<\q2<1000$~\g2} \\
\hline
$8-10  $ & $8.65  $ & $\pm 0.26  $ & $_{-0.24}^{+0.25}    $ & $_{-0.16}^{+0.14}    $ & $0.96$ & $0.89$ \\
$10-14 $ & $4.49  $ & $\pm 0.14  $ & $_{-0.11}^{+0.12}    $ & $_{-0.12}^{+0.13}    $ & $0.94$ & $0.88$ \\
$14-18 $ & $2.112 $ & $\pm 0.091 $ & $_{-0.073}^{+0.074}  $ & $_{-0.067}^{+0.073}  $ & $0.94$ & $0.90$ \\
$18-25 $ & $0.677 $ & $\pm 0.038 $ & $_{-0.024}^{+0.024}  $ & $_{-0.026}^{+0.031}  $ & $0.97$ & $0.89$ \\
$25-100$ & $0.0289$ & $\pm 0.0025$ & $_{-0.0010}^{+0.0010}$ & $_{-0.0018}^{+0.0016}$ & $0.93$ & $0.89$ \\
\hline
\multicolumn{7}{c}{$1000<\q2<2000$~\g2} \\
\hline
$8-10  $ & $3.30  $ & $\pm 0.16  $ & $_{-0.11}^{+0.12}    $ & $_{-0.06}^{+0.07}    $ & $0.92$ & $0.94$ \\
$10-14 $ & $2.024 $ & $\pm 0.096 $ & $_{-0.085}^{+0.080}  $ & $_{-0.036}^{+0.038}  $ & $0.95$ & $0.92$ \\
$14-18 $ & $1.070 $ & $\pm 0.068 $ & $_{-0.048}^{+0.049}  $ & $_{-0.025}^{+0.024}  $ & $0.96$ & $0.91$ \\
$18-25 $ & $0.478 $ & $\pm 0.033 $ & $_{-0.025}^{+0.026}  $ & $_{-0.014}^{+0.010}  $ & $0.93$ & $0.93$ \\
$25-100$ & $0.0246$ & $\pm 0.0024$ & $_{-0.0018}^{+0.0018}$ & $_{-0.0012}^{+0.0017}$ & $0.98$ & $0.93$ \\
\hline
\multicolumn{7}{c}{$2000<\q2<5000$~\g2} \\
\hline
$8-10  $ & $1.47  $ & $\pm 0.11  $ & $_{-0.05}^{+0.07}    $ & $_{-0.04}^{+0.00}    $ & $0.94$ & $0.86$ \\
$10-14 $ & $0.822 $ & $\pm 0.059 $ & $_{-0.031}^{+0.037}  $ & $_{-0.010}^{+0.020}  $ & $0.94$ & $0.94$ \\
$14-18 $ & $0.563 $ & $\pm 0.049 $ & $_{-0.034}^{+0.032}  $ & $_{-0.006}^{+0.012}  $ & $0.97$ & $0.92$ \\
$18-25 $ & $0.276 $ & $\pm 0.027 $ & $_{-0.015}^{+0.010}  $ & $_{-0.012}^{+0.006}  $ & $0.98$ & $0.99$ \\
$25-100$ & $0.0186$ & $\pm 0.0021$ & $_{-0.0008}^{+0.0008}$ & $_{-0.0005}^{+0.0006}$ & $0.88$ & $0.98$ \\
\hline
\multicolumn{7}{c}{$5000<\q2<100000$~\g2} \\
\hline
$8-10$ & $0.264$ & $\pm 0.044$ & $_{-0.052}^{+0.047}$ & $_{-0.001}^{+0.014}$ & $0.94$ & $0.90$ \\
$10-14$ & $0.248$ & $\pm 0.036$ & $_{-0.024}^{+0.019}$ & $_{-0.006}^{+0.011}$ & $0.96$ & $0.90$ \\
$14-18$ & $0.146$ & $\pm 0.027$ & $_{-0.012}^{+0.011}$ & $_{-0.002}^{+0.003}$ & $0.99$ & $0.91$ \\
$18-25$ & $0.082$ & $\pm 0.016$ & $_{-0.016}^{+0.015}$ & $_{-0.002}^{+0.002}$ & $1.00$ & $0.92$ \\
$25-100$ & $0.0111$ & $\pm 0.0021$ & $_{-0.0008}^{+0.0009}$ & $_{-0.0004}^{+0.0004}$ & $1.00$ & $0.99$ \\
\hline
    \end{tabular}
 \caption{
   The measured differential cross-section $\setjb$ for inclusive-jet
   production in different regions of $\q2$ using the SIScone jet
   algorithm. Other details as in the caption to Table~\ref{tabone}.}
 \label{tabfour}
\end{center}
\end{table}

\newpage
\clearpage

\newpage
\clearpage

\newpage
\clearpage

\newpage
\clearpage

\newpage
\clearpage

\newpage
\clearpage

\end{document}